\documentclass[a4paper,12pt]{article}
\usepackage{graphicx}
\usepackage{amsmath}
\usepackage{multirow}
\usepackage{float} 
\usepackage{url,tabularx,amsfonts,slashed,amsmath,array}
\usepackage{lscape}

\newcommand{\be}{\begin{eqnarray}}
\newcommand{\ee}{\end{eqnarray}}

\begin{document}

\begin{flushright}
preprint SHEP-11-12\\
preprint FR-PHENO-2012-002\\
\today
\end{flushright}
\vspace*{1.0truecm}

\begin{center}
{\large\bf $Z'$ signals in polarised top-antitop final states}\\
\vspace*{1.0truecm}
{\large L. Basso$^1$, K. Mimasu$^2$ and S. Moretti$^{2,3}$}\\
\vspace*{0.5truecm}
{\it $^1$Physikalisches Institut, Albert-Ludwigs-Universit\"at Freiburg\\
D-79104 Freiburg, Germany}\\
\vspace*{0.25truecm}
{\it $^2$School of Physics \& Astronomy, University of Southampton, \\
Highfield, Southampton, SO17 1BJ, UK}\\
\vspace*{0.25truecm}
{\it $^3$Particle Physics Department\\ Rutherford Appleton Laboratory\\
Chilton, Didcot, Oxon OX11 0QX, UK}\\
\end{center}

\vspace*{1.0truecm}
\begin{center}
\begin{abstract}
\noindent 
We study the sensitivity of top-antitop samples produced at all energy stages of the 
Large Hadron Collider (LHC) to the nature of an underlying $Z'$ boson, in presence of full tree level
standard model (SM) background effects and relative interferences. We concentrate on differential mass spectra as
well as both spatial and spin asymmetries thereby demonstrating that exploiting combinations of these
observables will enable one to distinguish between sequential $Z'$s and those pertaining to
Left-Right symmetric models as well as $E_6$ inspired ones, assuming 
realistic final state reconstruction efficiencies and error estimates.  
\end{abstract}
\end{center}

\section{Introduction}
\label{sect:intro}

A common feature of  $U(1)$ gauge extensions of the SM are $Z'$ bosons.
They appear for example in Little Higgs models~\cite{Schmaltz:2005ky}, 
in theories with extra spatial dimensions~\cite{Hewett:2002hv} as well as  
Grand Unified Theories (GUTs) such as $SO(10)$~\cite{mohapatra} (including Left-Right symmetric models) and 
$E_6$~\cite{Hewett:1988xc}. Further, in 
some Supersymmetry (SUSY) breaking scenarios~\cite{Chung:2003fi} and in Hidden Valley 
models~\cite{Strassler:2006im}, they can appear as messengers between the 
SM and the hidden sectors corresponding to such scenarios.

Such $Z'$ states are best searched for at hadron colliders through a di-lepton signature in the
neutral Drell-Yan (DY) process, i.e.,
$pp(\bar p) \to (\gamma,Z,Z') \to \ell^+\ell^-$, where $\ell=e,\mu$. In fact, the most stringent limits on $Z'$'s
at both Tevatron and the LHC come from this signature. 
The Tevatron (after all its runs) places limits on the $Z'$ mass, $M_{Z'}$, at around 1 TeV (for a sequential $Z'$) 
\cite{Tevatron} while the LHC (after the 7 TeV run) sets about 1.94 TeV (for the same $Z'$) \cite{LHC}\footnote{A sequential $Z'$ is a state with
generic mass and same coupling to the SM particles as the $Z$ boson. Limits in the aforementioned models are normally obtained
by rescaling the results for a sequential $Z'$, though this implicitly assumes that the $Z'$ cannot decay into
additional extra matters present in the model spectrum.}.
Since such an experimental signature is clean and theoretical uncertainties for inclusive quantities are well under control,
see, e.g., \cite{Fuks:2007gk}, including those associated to higher order effects,  both
(two-loop) QCD \cite{DYrefs} and (one-loop) EW \cite{Baur:2001ze} ones, one can conceive accessing the couplings of a discovered $Z'$ 
(thereby providing a window on its high scale genesis) by studying the ensuing di-lepton observables
(such as the leptonic invariant mass, angular distributions or spatial/spin asymmetries). Several phenomenological studies on how to 
measure $Z'$ properties and couplings to SM particles in the DY channel have been performed, see \cite{zprevs} for general studies and 
\cite{zcurr,Accomando:2010fz} for (rather recent) studies dedicated to specific models. 

Another decay channel of $Z'$ bosons carries phenomenological importance, despite being of less scope than the DY one in `discovering' such states,
i.e., $pp(\bar p)\to (\gamma,Z,Z') \to t\bar t$, into top-antitop quark pairs. Its reduced importance for $Z'$ searches with respect to the DY case 
is due to the much larger background (which includes QCD production in top-antitop quark pairs), the more involved final state yielding six or more objects
in the detector including jets and the associated poor efficiency in reconstructing the two heavy quarks. However, owing to the fact that the top (anti)quark
decays before hadronising (so that its spin properties are effectively transmitted to the decay products) and the electromagnetic charge of the top can be 
tagged (via a lepton and/or a $b$-jet) \cite{tt-pol}, $t\bar t$ samples can also be useful in profiling the $Z'$, 
as the aforementioned spatial/spin asymmetries (particularly 
effective
to pin down the couplings of the new gauge boson) can be defined in this case too \cite{Zp-tt}. 
Further, the large (anti)top mass induces non-trivial space/spin effects, which are not present in the DY case 
and are also sensitive to the nature of the intervening $Z'$ state.  
Experimental collaborations at both Tevatron \cite{Tevatron-tt}
and LHC \cite{LHC-tt} have in fact been pursuing this, in part driven by some anomalies emerged in the forward-backward asymmetry of $t\bar t$ samples
at the Tevatron \cite{anomaly}. Finally, just like in the case of DY, also for $t\bar t$ production higher-order effects from both QCD 
\cite{QCD-SM-tt} (see also \cite{earlycalc-QCD}) 
and EW \cite{EW-SM-tt} (see also \cite{earlycalc-EW}) interactions are well known, including the case of polarised (anti)tops.

The purpose of this paper is to study the sensitivity of the LHC (at all its planned energy stages) to the presence of a $Z'$ boson we well as
to assess the machine ability to profile a $Z'$ boson mediating $t\bar t$ production,
in both standard kinematic variables as well as spatial/spin asymmetries, by adopting some benchmark scenarios for
several realisations of the recalled sequential, Left-Right symmetric and $E_6$ based $Z'$ models (specifically, the same as those in
\cite{Accomando:2010fz}). The plan of the paper is as follow. In the next section we  describe
our calculation and define the observables to be studied. In sect.~\ref{sec:results} we report and comment on our results. Sect.~\ref{sec:summary} 
finally presents our conclusions.

\section{Calculation \label{sec:calculation}}

The code exploited for our study is based on helicity amplitudes, defined through the HELAS 
subroutines~\cite{HELAS}, and built up by means of MadGraph~\cite{MadGraph}. Initial state quarks have been taken
as massless while for the (anti)top state we have taken $m_t=175$ GeV as mass. The latter has been kept 
on-shell. The Parton Distribution Functions (PDFs) used were CTEQ6L1~\cite{cteq}, with factorisation/renormalisation
scale set to $Q=\mu=2m_t$. VEGAS~\cite{VEGAS} was used for the multi-dimensional numerical integrations.

The rest of this section introduces the asymmetry variables considered in this study of $Z'$ models in $t\bar{t}$ 
final states. We will eventually explore the effect of these observables aiming to define the most suitable ones for model discrimination purposes. 
We will define such quantities over the entire invariant mass spectrum of the $t\bar t$ system, though we will exploit these in differential form as well, i.e.,
as a function of $M_{t\bar t}$.

    \subsection{Spatial asymmetries}\label{subsec:chargedef}
		Charge or spatial asymmetry in collider physics is a measure of the symmetry of the angular dependence of the 
matrix element for the production of a two body final state. At a hadron collider, for example, defining a 
polar angle, $\theta$, as the angle between a final state particle and one of the incoming partons in the 
Centre-of-Mass (CM) frame, variables that are function of this angle can be constructed to probe the asymmetry of 
the distribution of said angle. Such an asymmetry can only be generated from a charge asymmetric initial state 
such as $q\bar{q}$, as opposed to, for example, the $gg$ initial state. In QCD, the asymmetry for the 
$t\bar{t}$ final state is generated dominantly at Next-to-Leading Order (NLO) via interference of leading order 
$q\bar{q}\to t\bar{t}$ with the corresponding box diagram as well as by the interference between 
initial and final state gluon radiation~\cite{QCDasymmetry}.
This will also have a comparable contribution from Electro-Weak (EW) processes, both at tree- and loop-level. 
        \subsubsection{Tevatron}
	Although not relevant for this paper (namely because no $Z'$ boson has been observed at the Tevatron), we introduce here the charge asymmetry studied at the Tevatron, mostly for clarity and completeness.
	
			The Tevatron, being a $p\bar{p}$ collider, is an ideal place to measure spatial asymmetries since 
the polar angle in the collider frame can more or less be identified with that of the CM frame modulo 
PDF effects. Statistically, both incoming partons will be valence quarks and an absolute preferred direction can 
be unambiguously defined. The forward-backward asymmetry $A_{FB}$, a possible measure of the aforementioned charge asymmetry,
 can then simply be defined by an integrated quantity as
	        \begin{align}\label{asy_A_FB_tevatron}
	            A_{FB}=\frac{N_{t(\bar{t})}(y>0)-N_{t(\bar{t})}(y<0)}{N_{t(\bar{t})}(y>0)+N_{t(\bar{t})}(y<0)},
	        \end{align} 
	   		where $y$ is the rapidity of the observed top (anti)quark and $N_{t(\bar{t})}$ denotes the 
number of top(antitop) quarks observed in the forward ($y>0$) or backward ($y<0$) direction. 
For this observable, the SM prediction is of order $5\%$, and, interestingly, both the CDF and D$\emptyset$ 
collaborations report a deviation~\cite{anomaly}\footnote{It is not the aim of this paper to address such a discrepancy, 
since models based on colorless bosons are virtually ruled out as a possible explanation~\cite{AguilarSaavedra:2011}.}.

\subsubsection{LHC}                  
The definition of a charge asymmetry at the LHC becomes somewhat more involved. First, the 
charge-symmetric di-gluon initial state dominates for a significant amount of the invariant mass range for 
$t\bar{t}$ final states, until the parton luminosities for $q\bar{q}$ become dominant, 
matching that of $gg$ at around $1\div 2$~TeV (depending on collider energy), and remaining 
significant from there on. This large zero contribution dilutes the predicted SM asymmetry down to $\sim 1\%$. Secondly, even when diagrams with initial state quarks become important, 
the fact that the $pp$ initial state is $\it C$-invariant necessitates the redefinition of the measured 
quantity itself. In this case, no preferred direction can be defined because the incoming 
quark will generally be a valence quark, while the antiquark must come from the sea. However, one can exploit the 
fact that the incoming quark will statistically carry a larger momentum fraction than the antiquark, resulting in 
a correlation between the boost of the $t\bar{t}$ frame and the direction of the incoming quark. Thus, 
defining the polar angle as, e.g., the angle between the incoming quark and the outgoing top quark, one expects a 
broadening of the top rapidity distribution with respect to the antitop one. The asymmetry can then be measured 
by restricting the rapidity range over which the top quarks are selected~\cite{Antunano:2007da}, constructing the so-called ``central'' ($C$) 
and ``forward'' ($F$) asymmetries:
	        \begin{align}\label{asy_AC}
	            A_{C}=\frac{N_{t}(|y|<y_{cut}^{C})-N_{\bar{t}}(|y|<y_{cut}^{C})}{N_{t}(|y|<y_{cut}^{C})+N_{\bar{t}}(|y|<y_{cut}^{C})},
	        \end{align}
	        \begin{align}
	            A_{F}=\frac{N_{t}(|y|>y_{cut}^{F})-N_{\bar{t}}(|y|>y_{cut}^{F})}{N_{t}(|y|>y_{cut}^{F})+N_{\bar{t}}(|y|>y_{cut}^{F})}.
	        \end{align}
			Here, a rapidity cut, $y_{cut}^{C(F)}$, is chosen and the number of tops ($N_{t}$) and of antitops ($N_{\bar{t}}$) in that region are compared. Taking $y_{cut}^{C(F)}\rightarrow\infty(0)$, for $A_{C}$($A_{F}$), respectively (i.e., integrating them over the whole 
rapidity range), will restore $A_{C(F)}=0$. In this analysis, $y_{cut}^{C}=y_{cut}^{F}=0.5$.
			
			Alternatively, one can construct variables that are function of the absolute rapidity difference of the top and antitop quarks, $\Delta y=|y_{t}-y_{\bar{t}}|$, which is sensitive to the CM polar angle and independent of the $t\bar{t}$ frame boost. We implemented this in 
two ways depending on asymmetry-enhancing kinematical cuts proposed in~\cite{Zhou:2011dg}, acting on the rapidity or on the 
momentum along the beam axis (e.g., the $z$ one) of the top-antitop pair, $y_{t\bar{t}}=\tfrac{1}{2}(y_{t}+y_{\bar{t}})$ and $p^{z}_{t\bar{t}}=p^{z}_{t}+p^{z}_{\bar{t}}$,
			 referred to as ``rapidity dependent'' ($RFB$) and ``one-sided''  ($OFB$) forward-backward asymmetries, respectively:
            \begin{align}
                A_{RFB}&=\frac{N(\Delta y > 0)-N(\Delta y < 0)}{N(\Delta y > 0)+N(\Delta y < 0)}\Bigg |_{|y_{t\bar{t}}|>|y^{cut}_{t\bar{t}}},
            \end{align}
            \begin{align}
                A_{OFB}&=\frac{N(\Delta y > 0)-N(\Delta y < 0)}{N(\Delta y > 0)+N(\Delta y < 0)}\Bigg |_{|p^{z}_{t\bar{t}}|>p^{z}_{cut}}.
            \end{align}
Such kinematical cuts are designed to enhance the contributions from the $q\bar{q}$ initial state by probing regions of high partonic momentum fraction, $x$, where the parton luminosity of interest is more important. 			
    \subsection{Spin asymmetries}\label{subsec:spindef}
		Since the top decays before it hadronises, the helicity information is preserved in its 
decay channels. Two relevant spin asymmetries, one double ($LL$) and one single ($L$), can then be defined as follows:
	        \begin{align}\label{asy_ALL}
	            A_{LL}=\frac{N(+,+) + N(-,-) - N(+,-) - N(-,+)}{N_{Total}}\quad;
	        \end{align}
	        \begin{align}\label{asy_AL}
	            A_{L}=\frac{N(-,-) + N(-,+) - N(+,+) - N(+,-)}{N_{Total}}\quad.
	        \end{align}
			Here $N$ denotes the number of observed events and its first(second) argument corresponds to 
the helicity of the top (anti)quark. Notice that such spin asymmetries can be equally defined at
both the Tevatron and LHC, as they only pertain to the final state. As previously stated, such quantities can only be measured experimentally for a leptonic or top quark final state.

Defining our generic neutral current interaction in terms of gauge, vector and axial couplings $g^{\prime}$, $g_{V}$ and $g_{A}$ with the Feynman rule
\begin{equation}
	i\frac{g^{\prime}}{2}\gamma^{\mu}(g_{V}-g_{A}\gamma^{5}),
\end{equation}
\noindent
the dependence on the chiral couplings of the spin asymmetries can be expressed analytically, using helicity formulas from Ref.~\cite{Arai:2008qa} (also derived independently with the guidance of~\cite{Hagiwara:1985yu}): 
\begin{eqnarray}\label{analytic_ALL}
A_{LL}^{i} &\propto& \Big( 3\, (g^t_A)^2\beta^2 + (g^t_V)^2(2+\beta^2)\Big) \, \Big( (g^i_V)^2 + (g^i_A)^2 \Big) \, ,\\ \label{analytic_AL}
A_{L}^{i}  &\propto& g^t_A\, g^t_V \,\beta\,\Big( (g^i_V)^2 + (g^i_A)^2 \Big) \, ,
\end{eqnarray}
for a neutral gauge boson exchanged in the $s$-channel. Here, $i$ labels the initial state partons, $\beta=\sqrt{1-4\, m_t^2/\hat{s}}$ ($\hat{s}$ being the partonic CM energy of the process) and the angular dependence has been integrated over. These imply that $A_{LL}$ depends only on the square of the couplings similarly to the total cross section and that $A_{L}$ is only non-vanishing for non-zero vector and axial couplings of the final state tops.

\label{sec:definitions}
\subsection{Benchmark models}
The present analysis employs some of the benchmark models of table~1 in Ref.~\cite{Accomando:2010fz}, from the $E(6)$, $G_{LR}$ and $G_{SM}$ parameterisations.
While a brief overview is presented here, we refer the reader to the original publication for further details.

In the $E(6)$ case, a $Z'$ boson naturally arises from the $E_6 \to SU(5) \times U(1)_\psi  \times U(1)_\chi$ pattern of symmetry breaking, where it is expected that a certain linear combination of the two $U(1)$ factors, parametrized by a mixing angle $\theta$, survives down to the TeV scale:  $U(1)'=\cos{\theta}\,U(1)_\chi + \sin{\theta}\,U(1)_\psi$ (applying similarly to the symmetry generator $Q_{E_6}$). The coupling to fermion reads $g'Q_{E_6}Z'$, where $g'= 0.462$.

The generalised Left-Right (GLR) model originates from a left-right symmetric model, in which a TeV scale $Z'$ boson arises, related to the following pattern of symmetry breaking: $U(1)_R \times U(1)_{B-L} \to U(1)_Y$. While in the original model only a specific combination of $U(1)_R$ and $U(1)_{B-L}$ survives, the generalised version allows a generic linear combination: $U(1)'=\cos{\phi}\,U(1)_R + \sin{\phi}\,U(1)_{B-L}$ (applying similarly to the symmetry generator $Q_{GLR}$). The coupling to fermions reads as $g'Q_{GLR}Z'$, where $g'= 0.595$.

The generalised sequential SM (GSM) is a class of models the includes the traditional sequential $Z'$ boson, or $Z'_{SSM}$, a heavier copy of the SM-$Z$ boson. Such a boson has identical couplings to the SM-$Z$, with $Q_Z=Q_{T^3_L} -s^2_W Q$, $s^2_W$ being the sine of the weak angle and $Q$ the electric charge generator, respectively. Similarly to the GLR class, the GSM class is defined as a generalisation of the $Z$ coupling mixing, such that $Q_{GSM}=\cos{\alpha}\,Q_{T^3_L} + \sin{\alpha}\,Q$, so that the $Z'$ coupling to fermions is $g'Q_{GSM}Z'$, where $g'=0.76$.

The benchmark models that are relevant for the present paper are collected in table~\ref{tab_benchmarks}, with the coupling to quarks split in vector and axial parts. Universal couplings between generations are assumed.

\begin{table}[!h]
\centering
\begin{tabular}{lccccc}
\hline
$U(1)'$ & Parameter & $g_V^u$ & $g_A^u$ & $g_V^d$ & $g_A^d$ \\
\hline
\hline
 $E_6$ $(g'=0.462)$  & $\theta$  &      &    &   &     \\
\hline
$U(1)_{\chi}$ & 0 & 0    &  -0.316  &  -0.632 & 0.316    \\
$U(1)_{\psi}$ & $0.5\pi$ & 0    &  0.408  &      0   & 0.408    \\
$U(1)_{\eta}$  & -$0.29\pi$ & 0    &  -0.516  &  -0.387 & -0.129   \\
$U(1)_S$  & $0.129\pi$ &    0    &  -0.129  &  -0.581 & 0.452  \\
$U(1)_N$   & $0.42\pi$ &    0    &  0.316  &  -0.158 & 0.474 \\
\hline
\hline
 $G_{LR}$  $(g'=0.595)$    & $\phi$ &      &    &   &    \\
\hline
$U(1)_{R}$  & 0 & 0.5    &  -0.5  &  -0.5 & 0.5  \\
$U(1)_{B-L}$  & $0.5\pi$ & 0.333    &  0  &  0.333 & 0  \\
$U(1)_{LR}$ & $-0.128\pi$& 0.329    &  -0.46  &  -0.591 & 0.46\\
$U(1)_{Y}$ & $0.25\pi$ & 0.589    &  -0.354  &  -0.118 & 0.354  \\
\hline
\hline
$G_{SM}$   $(g'=0.760)$ & $\alpha$   &      &    &   &    \\
\hline
$U(1)_{SM}$ & $-0.072\pi$ & 0.193    &  0.5  &  -0.347 & -0.5  \\
$U(1)_{T_{3L}}$ & $0$& 0.5    &  0.5  &  -0.5 & -0.5 \\
$U(1)_Q$ & $0.5\pi$& 1.333  &  0 & -0.666  & 0   \\
\hline
\hline
\end{tabular}
\caption{Benchmark model parameters and couplings. The angles $\theta$, $\phi$, 
$\alpha$ are defined in the text~\cite{Accomando:2010fz}.
\label{tab_benchmarks}}
\end{table}

The table shows that the $E_{6}$($G_{LR}({BL})$) models share a general feature of the $Z^{\prime}$ having a purely axial(vector) coupling 
to up-type quarks. Generating a tree-level charge asymmetry requires both non-zero vector and axial couplings, since the asymmetric term in the matrix element ($\propto\cos\theta$) is proportional to $g^{i}_{V}g^{i}_{A}g^{t}_{V}g^{t}_{A}$ ($i$ being the initial state fermion and $t$ the top quark), where the $g_{V,A}$'s are defined in 
section~\ref{sec:benchmarks}. Hence, charge asymmetry can only be generated via interference with the $\gamma,Z$ background and is therefore expected to be very small for these models. Similarly for the spin asymmetries, eqs~(\ref{analytic_ALL})--(\ref{analytic_AL}) imply that the $E_{6}$-type models will only have a non-vanishing $A_{LL}$ (at the $Z'$ boson peak), that could serve as an extra handle to pin down parameters for these models, while $A_L$ is, again, generated only via interference. 

The rest of the models have generic, non-zero vector and axial couplings which will generate charge asymmetry, $A_{LL}$ and $A_{L}$. As mentioned in section~\ref{subsec:spindef}, $A_{L}$ has the extra handle of distinguishing relative sign between the vector and axial couplings. Looking at the table, one would therefore expect the $G_{LR}$ and $G_{SM}$ subclasses to therefore have opposite signs providing a clear distinguishing feature.

For these reasons, we will subdivide our analysis into that of the `$E_{6}$' type models which will also include $G_{LR}({BL})$ and `Generalised' models comprising of the rest.
 \label{sec:benchmarks}
\section{Results \label{sec:results}}

We present here a selection of results profiling the spatial and spin asymmetry distributions of the 
benchmark $Z^{\prime}$ models, that were summarised in the 
previous section. 
The variables described in section~\ref{sec:calculation} were computed as a 
function of the $t\bar{t}$ invariant mass within $\Delta M_{t\bar t}=|M_{Z^{\prime}}-M_{t\overline{t}}|<500$ GeV and compared to the tree-level SM predictions\footnote{t has also been shown that the fraction of $q\bar{q}$ initiated events could be equally enhanced by other kinematical cuts, such as on the transverse momentum of each top quark~\cite{Hewett:2011wz} as well as on the $t\bar{t}$ system~\cite{Kuhn:2011ri,Alvarez:2012vq}. However, notice that the latter is not applicable to tree level studies.}.
$Z^{\prime}$ boson masses of $1.7$ and $2.0$ TeV were taken and simulated for the LHC at $7$, $8$ and $14$ TeV. Only results for $8$ and $14$ TeV CM energy are 
presented. However, the former energy can still be taken as representative for the $7$ TeV run, as the corresponding results are very similar.

To be able to quantitatively address the distinguishability among the various models and the SM background, the statistical error 
of the predictions was calculated for some specific integrated luminosities. Given that an asymmetry is defined in terms of the number of events measured in 
some generic `forward' ($N_{F}$) and `backward' ($N_{B}$) directions (this is also true for spin asymmetries), the statistical error is evaluated by 
propagating the Poisson error on each measured 
quantity (i.e., $\delta N_{F(B)}=\sqrt{N_{F(B)}}$). Per given integrated luminosity $\mathcal{L}$, the measured number of events will be $N=\varepsilon\mathcal{L}\sigma$, $\sigma$ being the cross section, yielding an uncertainty on the asymmetry $A$ of 
\begin{equation}\label{eqn:error}
	\delta A\equiv \delta\left(\frac{N_{F}-N_{B}}{N_{F}+N_{B}}\right)=\sqrt{\frac{2}{\mathcal{L}\varepsilon}	
\left(\frac{\sigma^{2}_{F}+\sigma^{2}_{B}}{\sigma_{Total}^{3}}\right)}.
\end{equation}
 Here, $\varepsilon$  corresponds to an assumed $10\%$ reconstruction efficiency of the 
$t\bar{t}$ system, considering all possible decay $t\bar t$ decay channels~\cite{tdr}. The continuous curves on the following plots are the central values for the given asymmetry, with a statistical error quantified by binning the cross sections in 
$M_{t\overline{t}}$ for a bin width of $50$ GeV compatible with typical experimental resolutions in this quantity.
We also show a selection of two bin plots integrating the cross sections over an `on-peak' range ($\Delta M_{t\bar t}<100$ GeV) and evaluating the corresponding 
partially integrated asymmetry. Finally the totally integrated 
asymmetries are summarised in the corresponding tables, for both the $M_{t\overline{t}}$ window cuts. Invariant mass distributions of the total cross sections for $t\overline{t}$ production are also included for reference, with the statistical error normalised by the bin width.

Although in this work we only estimate statistical uncertainties,
systematics may be important as well~\cite{Hewett:2011wz,Alvarez:2012vq}. However, although the mass window selection is expected to milden their actual contribution, their inclusion would require detailed detector simulations which are beyond the scope of this paper. In this respect, it should further be noted that, by the time the LHC will reach the 14 TeV stage, where our most interesting results are applicable, systematics will be much better understood than at present.

Furthermore, the statistical significance of the measures (based on the assumption that they are independent) is calculated as follows: 
\begin{equation}\label{eqn:signif}
s \equiv \frac{\left| A(1)-A(2)\right|}{\sqrt{\delta A(1)^2 + \delta A(2)^2}}\, .
\end{equation}
We will make use of eq.~(\ref{eqn:signif}) to establish the disentanglement power of the LHC for $\sqrt{s}=14$ TeV with $100$ fb$^{-1}$. Finally, given that the statistical error of eq.~(\ref{eqn:error}) has a naive $1/\sqrt{\mathcal{L}}$ dependance, it is clear that the significance of a measure, as for eq.~(\ref{eqn:signif}), increases with luminosity. Inverting eq.~(\ref{eqn:signif}) and solving for $\mathcal{L}$, we also determine the required luminosity to distinguish the models with respect to the irreducible SM background and among themselves, defining disentanglement by $s \geq 3$.


In the following, we will first present and comment on the differential distributions for the most significant asymmetries in each class, i.e. $A_{LL}$ for the $E_6$-type models and $A_{LL}$, $A_L$ and $A_{RFB}$ for the generalised models. The comparison among the two classes (and between elements in them) is done at the end, evaluating the significance of the presented distributions.

\subsection{$E_{6}$-type models \label{subsec:E_6_results}}

\begin{figure}
\centering
\includegraphics[width=0.4\linewidth]{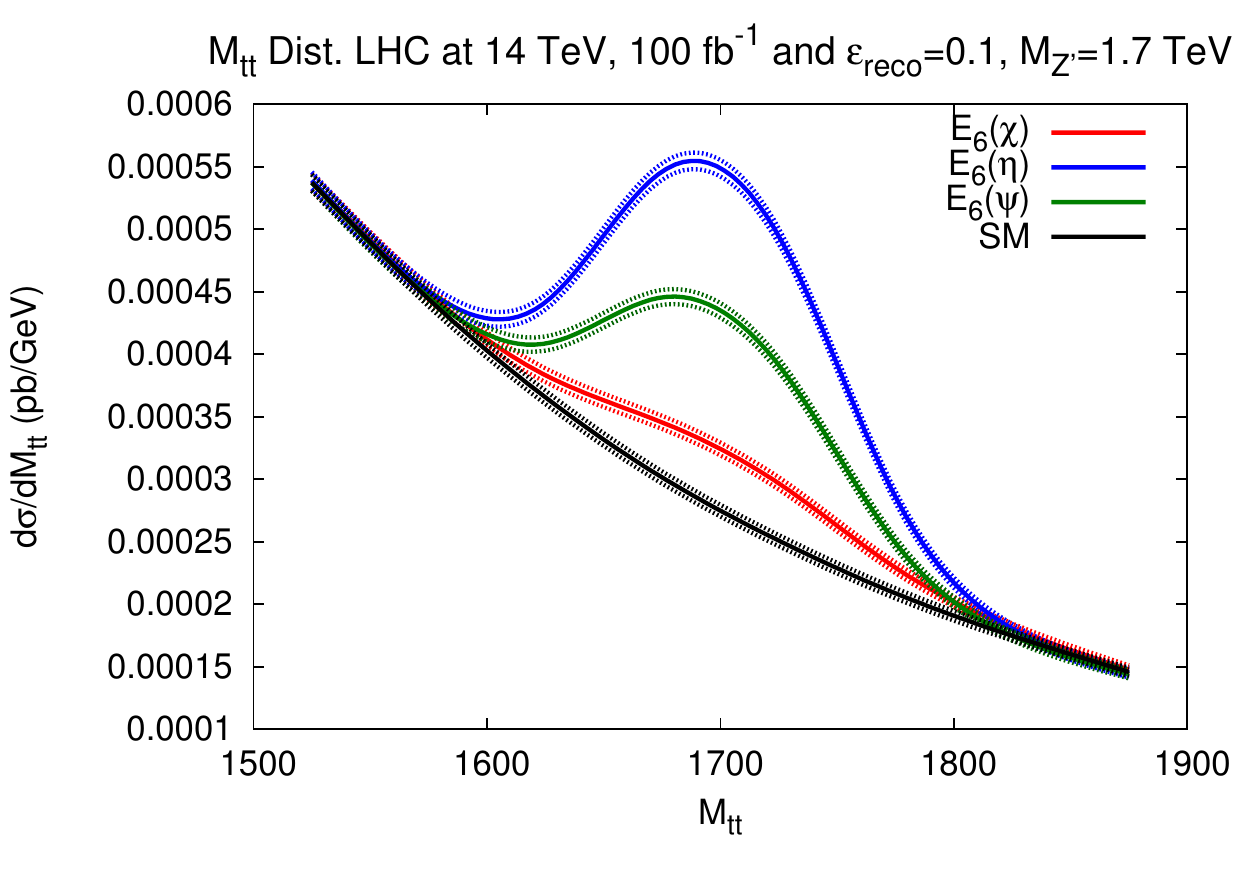}
\includegraphics[width=0.4\linewidth]{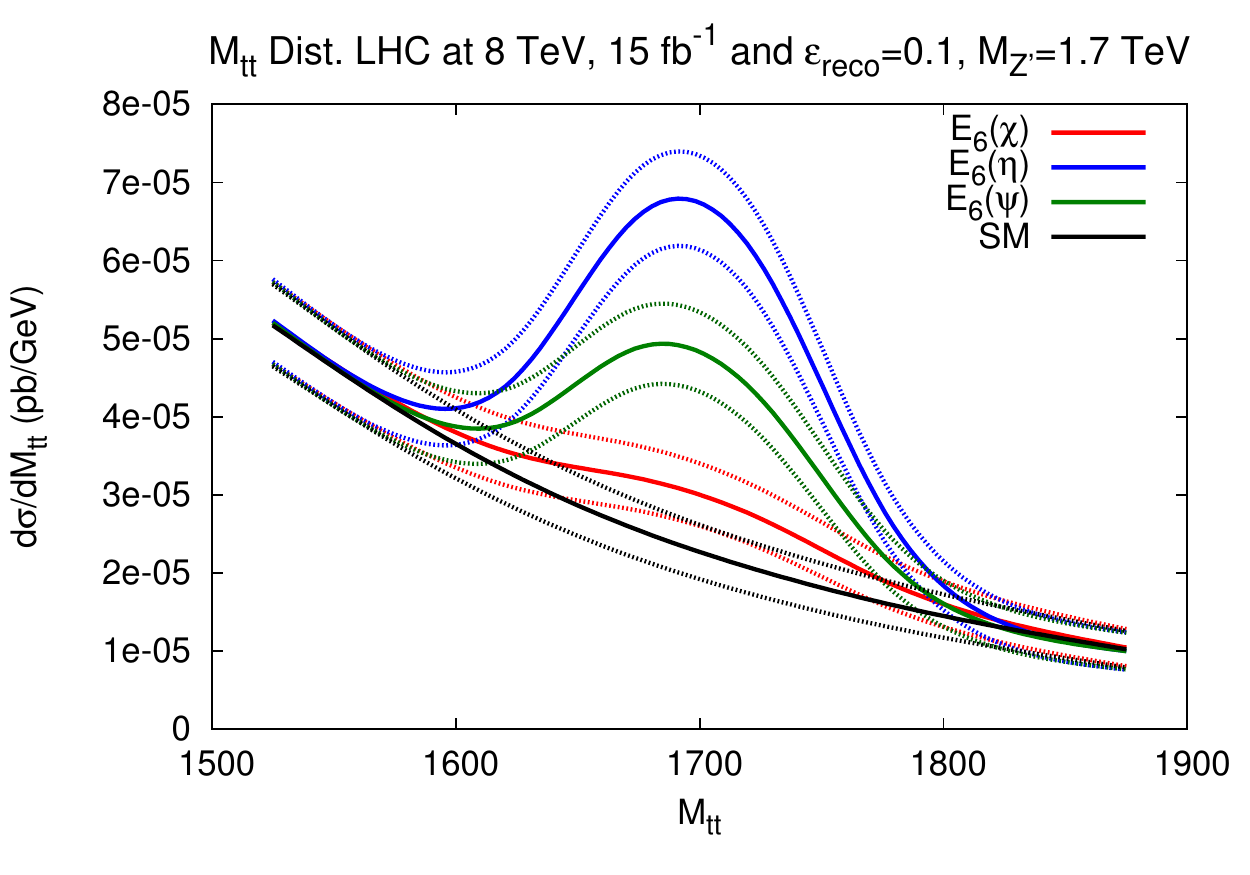}\\
\includegraphics[width=0.4\linewidth]{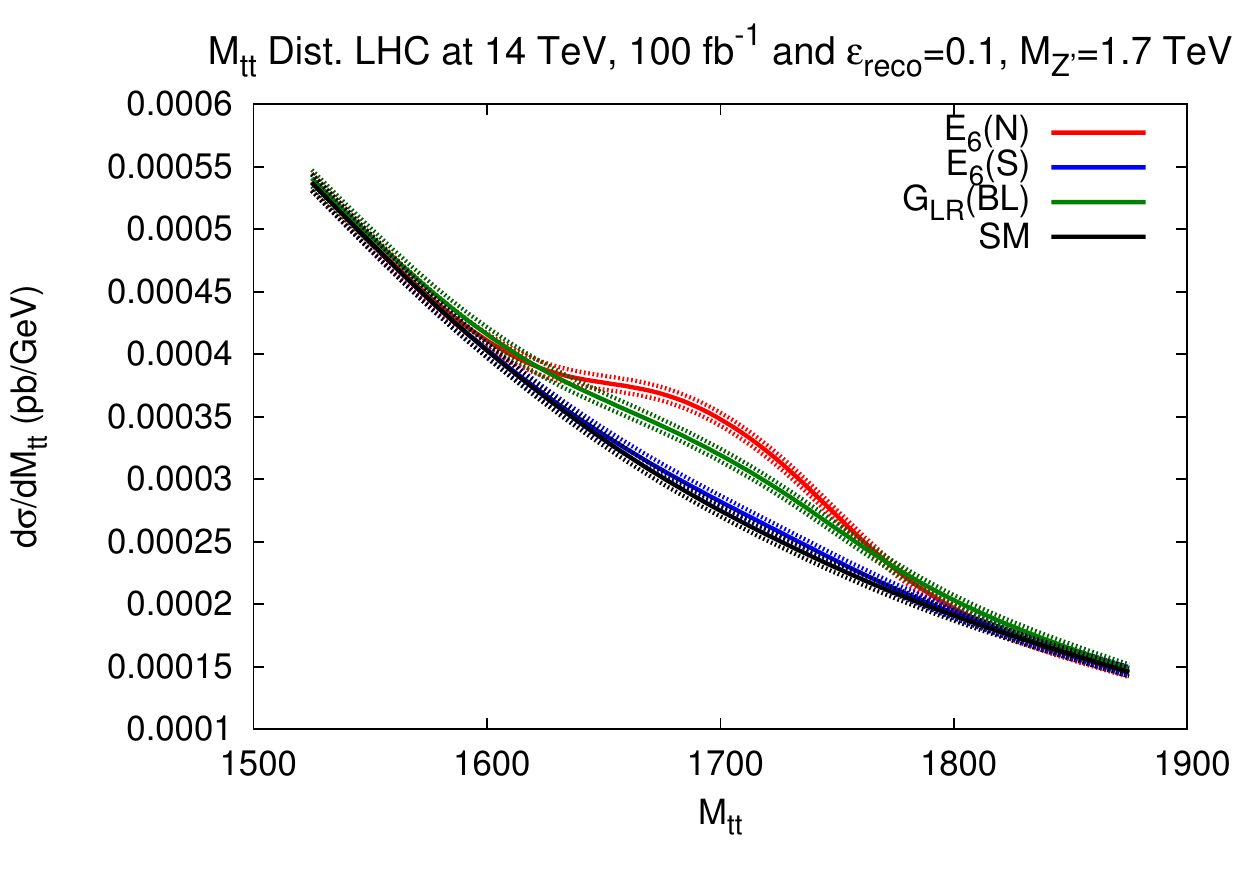}
\includegraphics[width=0.4\linewidth]{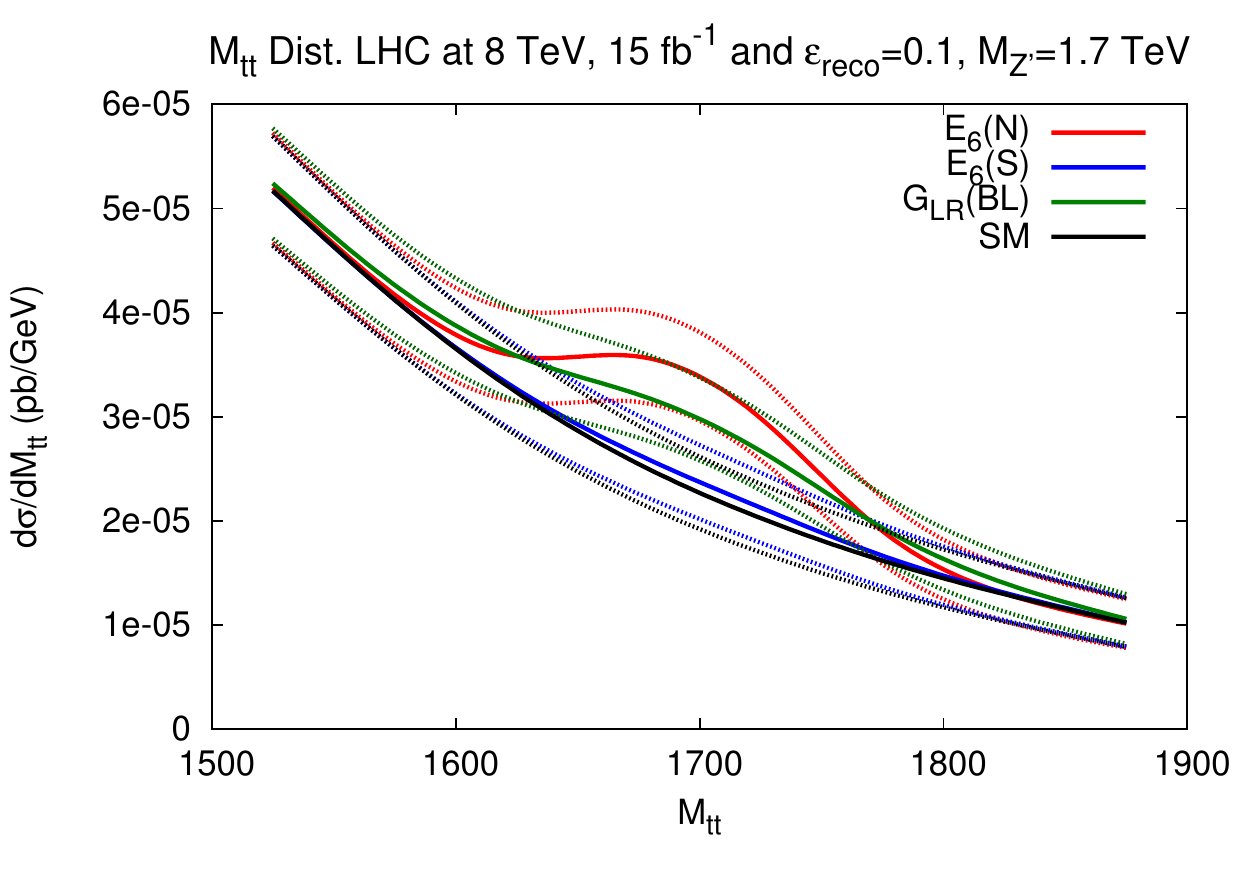}
\caption{Invariant mass distributions for $E_{6}$-type models for $M_{Z^{\prime}}$=1.7 TeV for the LHC at 14(8) TeV 
and 100(15)~fb$^{-1}$ of integrated luminosity.}\label{fig:LHC_E_6_Mtt}
\end{figure}

\begin{figure}[h!]	
	\centering
\includegraphics[width=0.32\linewidth]{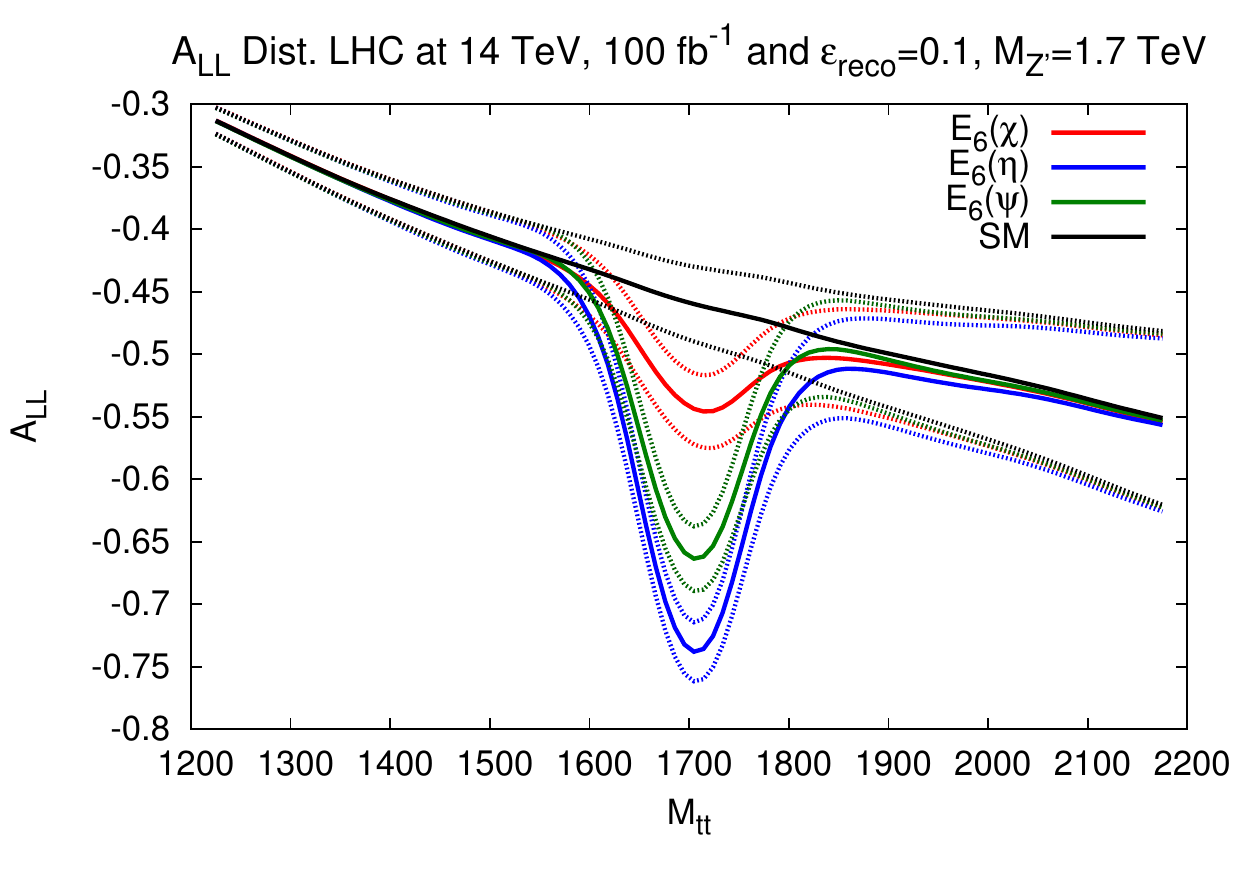}
\includegraphics[width=0.32\linewidth]{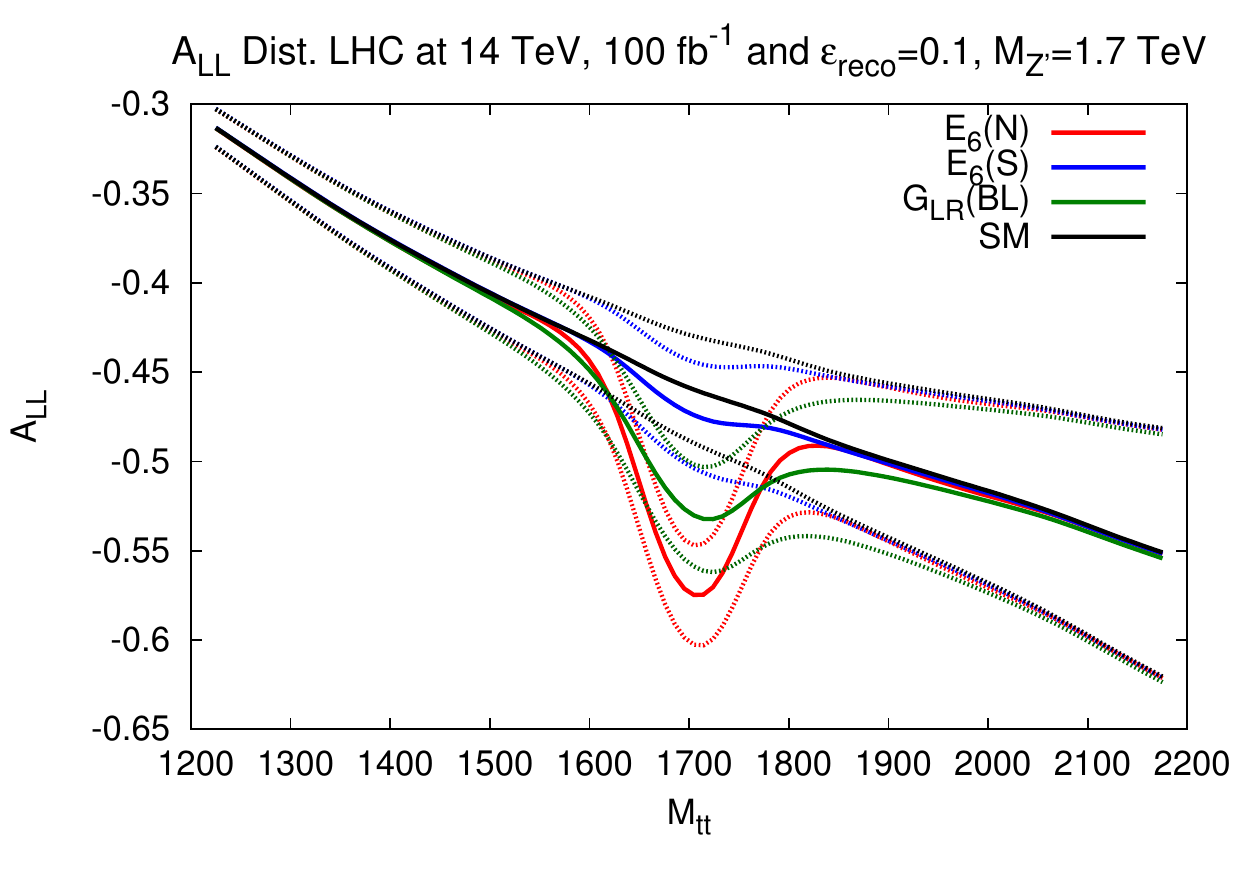}
\includegraphics[width=0.32\linewidth]{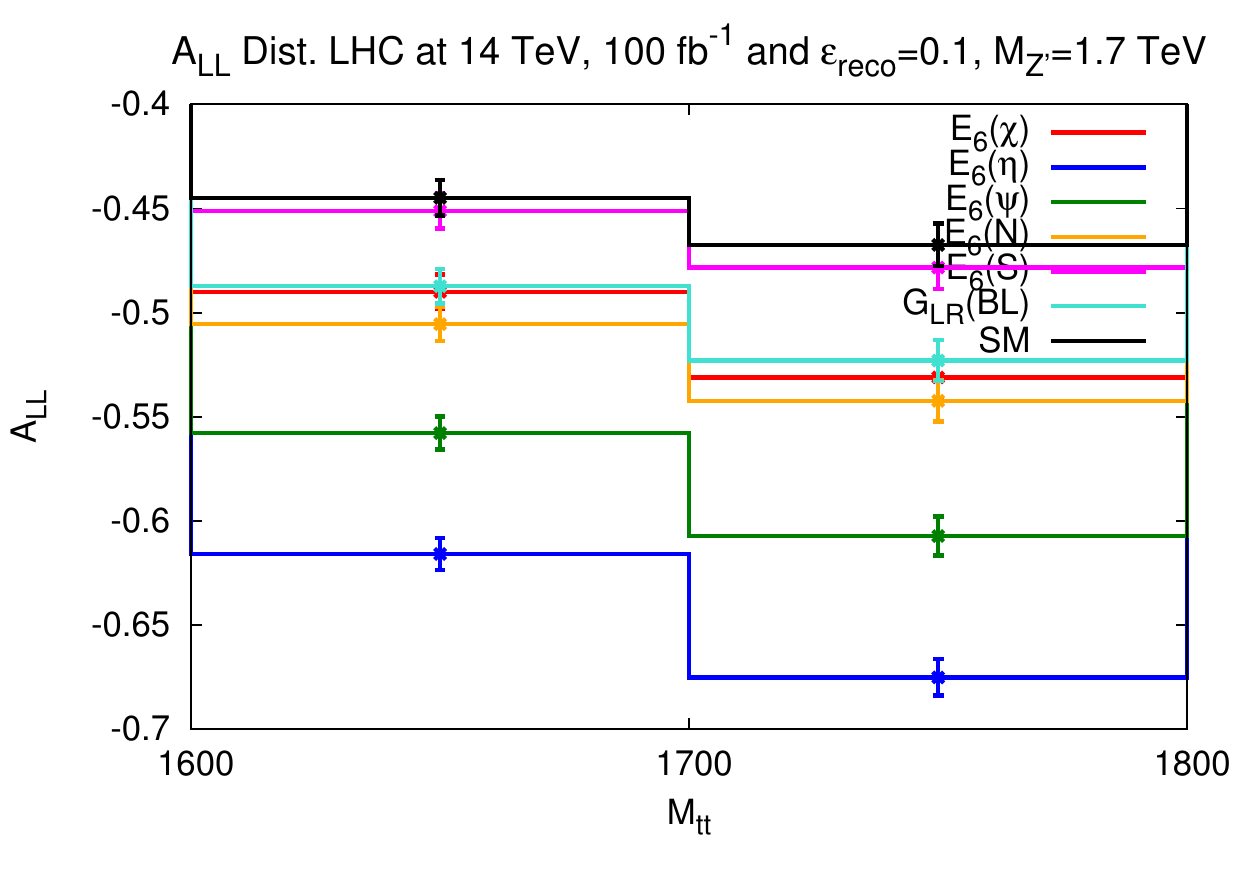}\\
\includegraphics[width=0.32\linewidth]{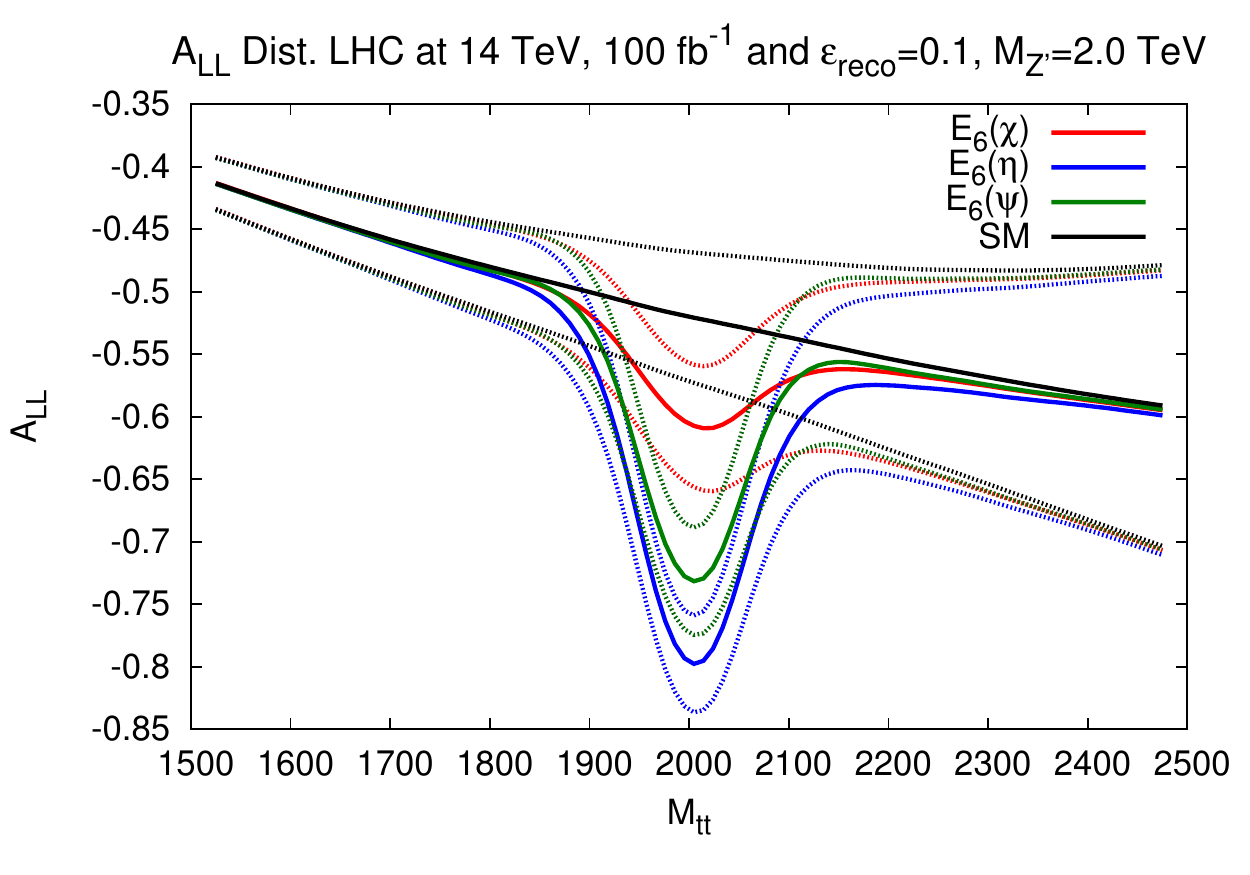}
\includegraphics[width=0.32\linewidth]{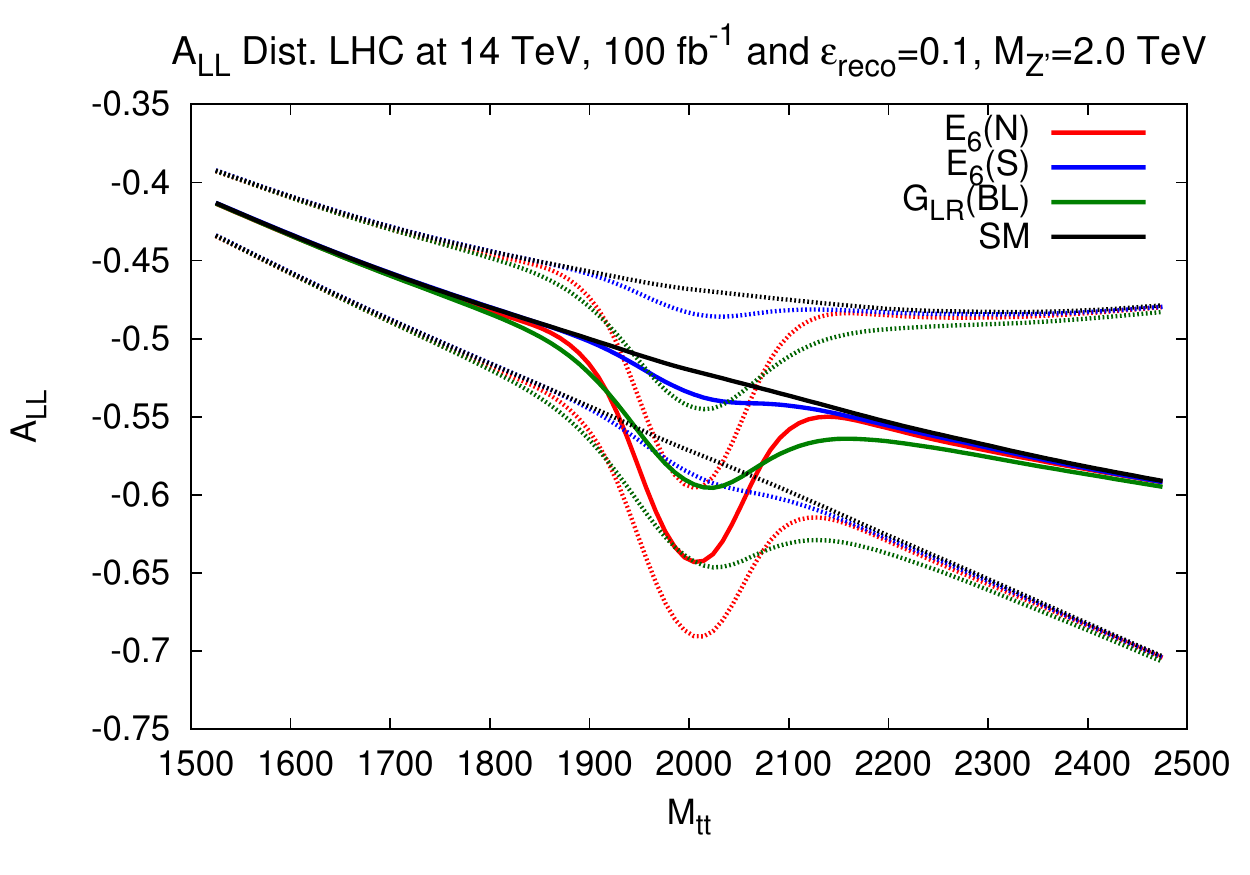}
\includegraphics[width=0.32\linewidth]{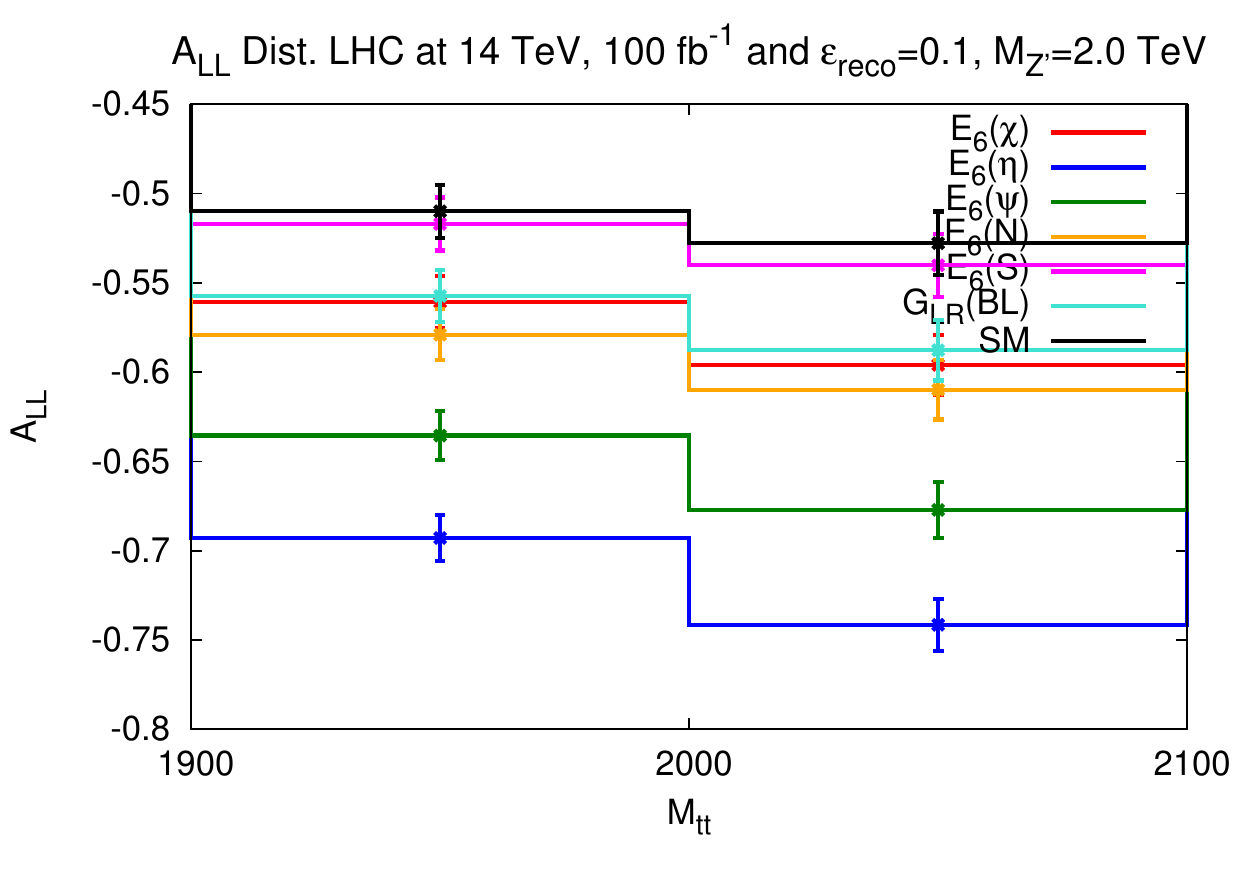}
\caption{$A_{LL}$ binned in $M_{t\overline{t}}$ for $E_{6}$-type models with $M_{Z^{\prime}}$=1.7 (\emph{upper}) and 2 (\emph{lower}) TeV
for the LHC at 14 TeV assuming 100 fb$^{-1}$ of integrated luminosity. Rightmost plots show the distribution in two 100 GeV bins either side of the 
$Z^{\prime}$ peak. Dotted lines and error bars represent statistical uncertainty calculated as described in the text.}\label{fig:LHC14_E_6_ALL}
\end{figure}

Figure~\ref{fig:LHC_E_6_Mtt} presents the invariant mass distributions around the 
$Z'$ peak for all models with $M_{Z^{\prime}}=$1.7 TeV at both 14 and 8 TeV. These plots show that the 
various $Z'$ bosons would certainly be visible in this 
channel, 
especially in the high energy and high luminosity scenario.
Figure~\ref{fig:LHC14_E_6_ALL} profiles $A_{LL}$ in invariant mass for the LHC at 14 TeV and Table~\ref{tab:LHC_E_6_ALL} summarises integrated values for these models at both 14 and 8 TeV. When 
calculating statistical uncertainties, an integrated luminosity of 100 and 15 fb$^{-1}$ is assumed for the two energies, respectively.

The analytic expression in eq.~(\ref{analytic_ALL}) shows that the observable depends on the top couplings in a similar way to the total cross section. This is reflected in the deviations from the SM shown in the figures, with large effects occurring on peak whose significances increase when restricting $\Delta M_{t\bar t}$ as shown in Table~\ref{tab:LHC_E_6_ALL}. This more or less parallels the effects seen in the invariant mass distributions. In the limit of $\hat{s}>>4m^{2}_{t}$, $A_{LL}$ depends identically on the vector and axial couplings of the tops and therefore cannot distinguish between the purely vector and purely axial cases of $G_{LR}(B-L)$ and $E_6$ models. Furthermore, Unlike $A_{L}$, it is insensitive to the relative sign between the couplings. These features are reflected by the 
overlapping of the $G_{LR}(B-L)$ and $E_{6}(\chi)$ cases, that differ in having purely vector and axial couplings respectively of different sign but of a similar magnitude. Such cases are never distinguishable neither with total cross section nor $A_{LL}$ measurements.

Aside from these limitations, Figure~\ref{fig:LHC14_E_6_ALL} shows clear distinguishability of models from the SM and between one another based on differences in couplings for the high energy case except when the up type coupling is too small, as for $E_{6}(S)$. (Table~\ref{tab_benchmarks} implies that this model would be much better suited to the $b\bar{b}$ channel.) Table~\ref{tab:LHC_E_6_ALL} further improves on these numbers by comparing integrated values focused around the $Z^{\prime}$ peak which gives scope for sensitivity to deviations from the SM and limited distinguishability even at low energy.

\subsection{Generalised models \label{subsec:GXX_results}}
\begin{figure}
	\centering
	\includegraphics[width=0.4\linewidth]{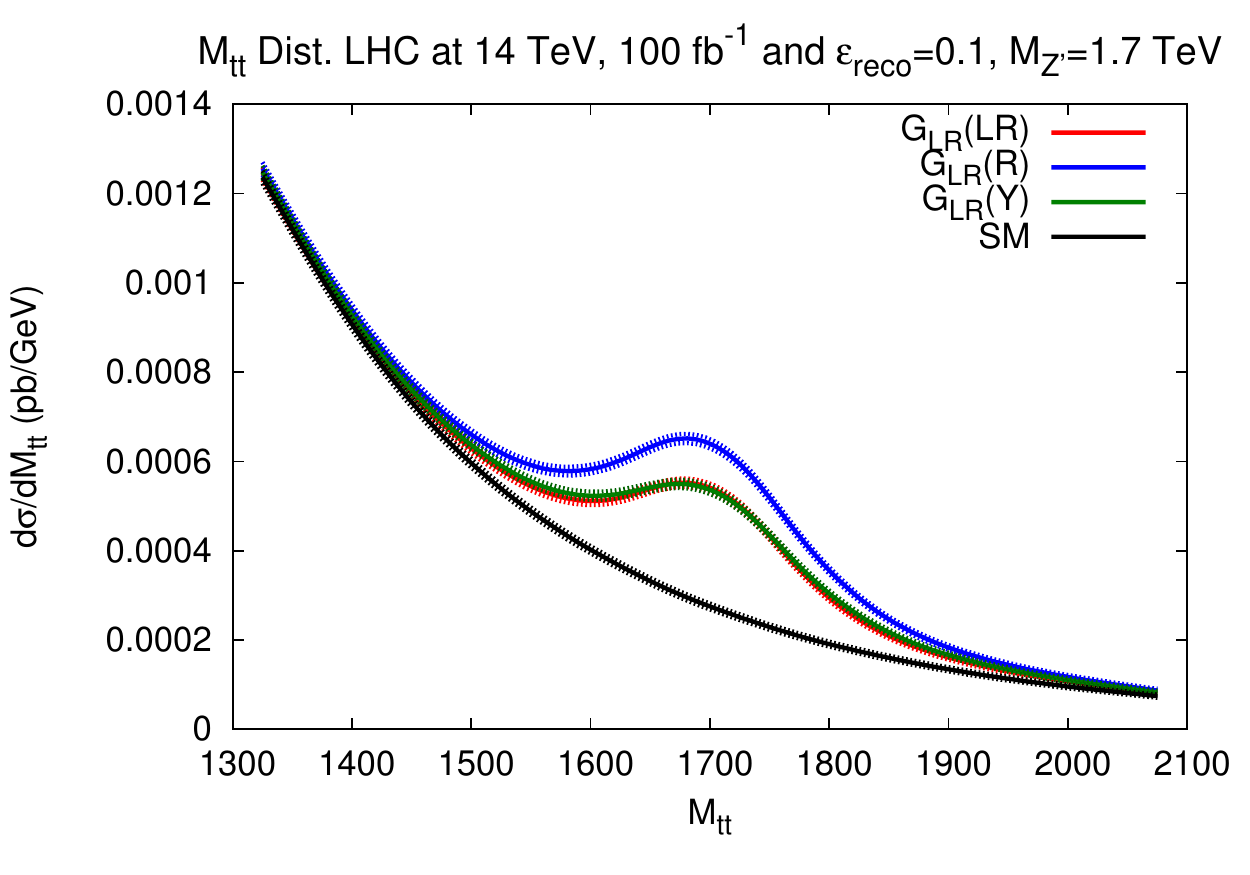}
	\includegraphics[width=0.4\linewidth]{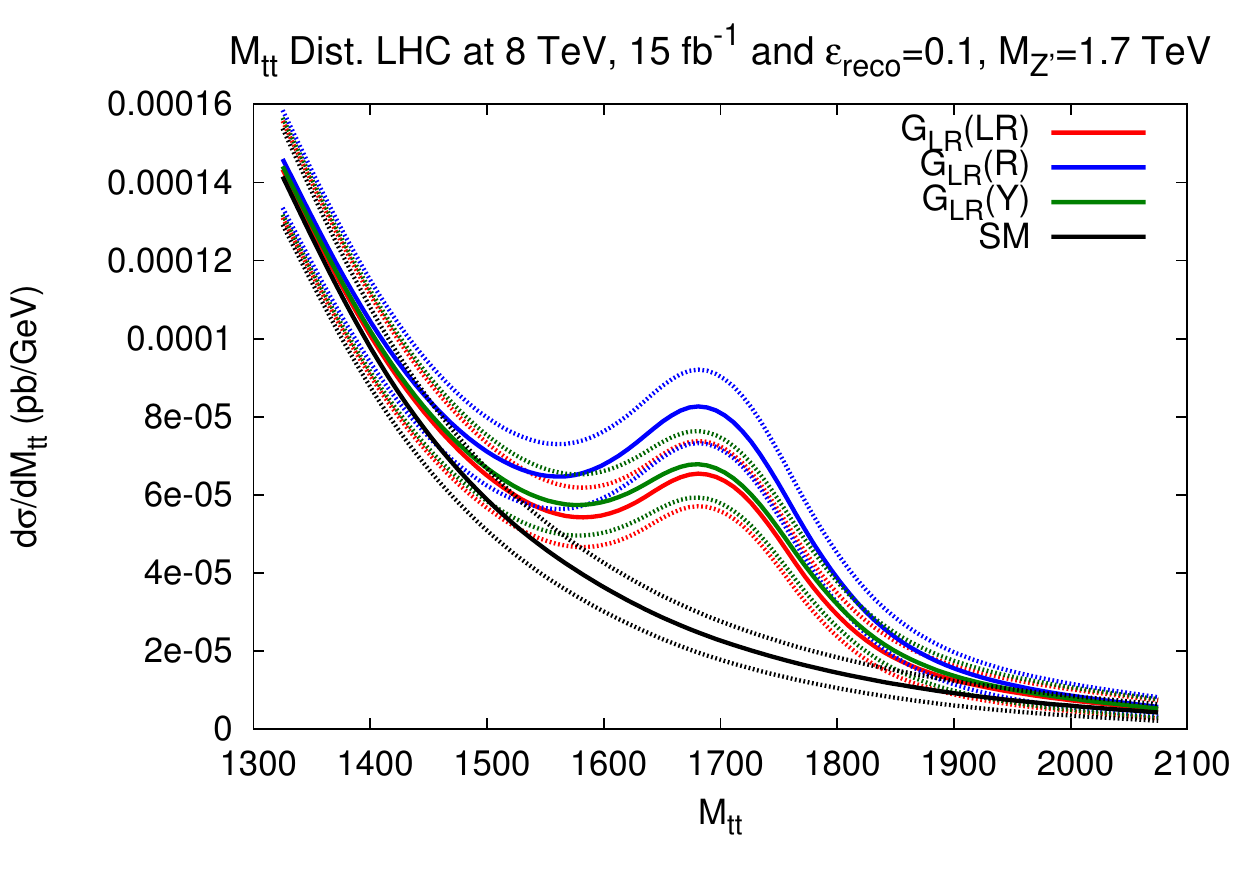}\\
	\includegraphics[width=0.4\linewidth]{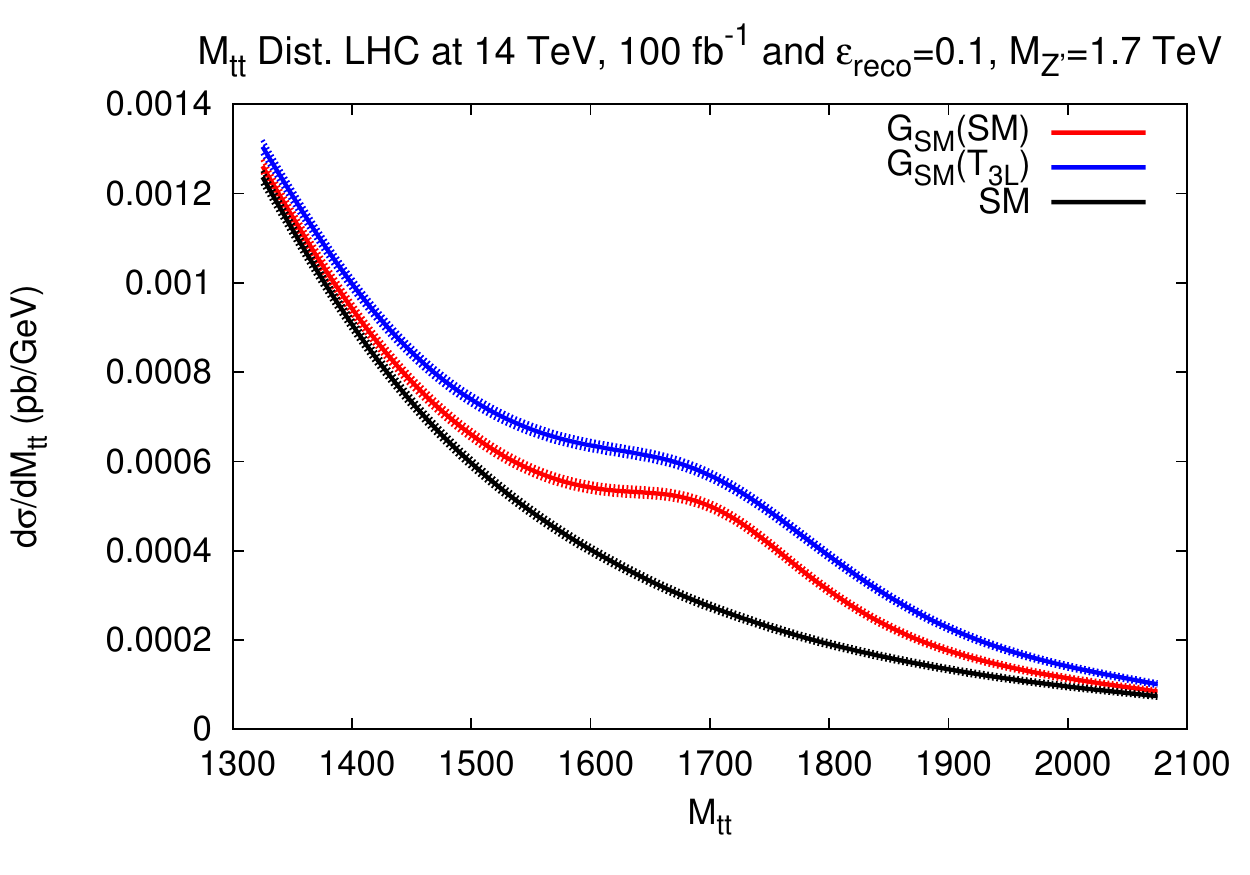}
	\includegraphics[width=0.4\linewidth]{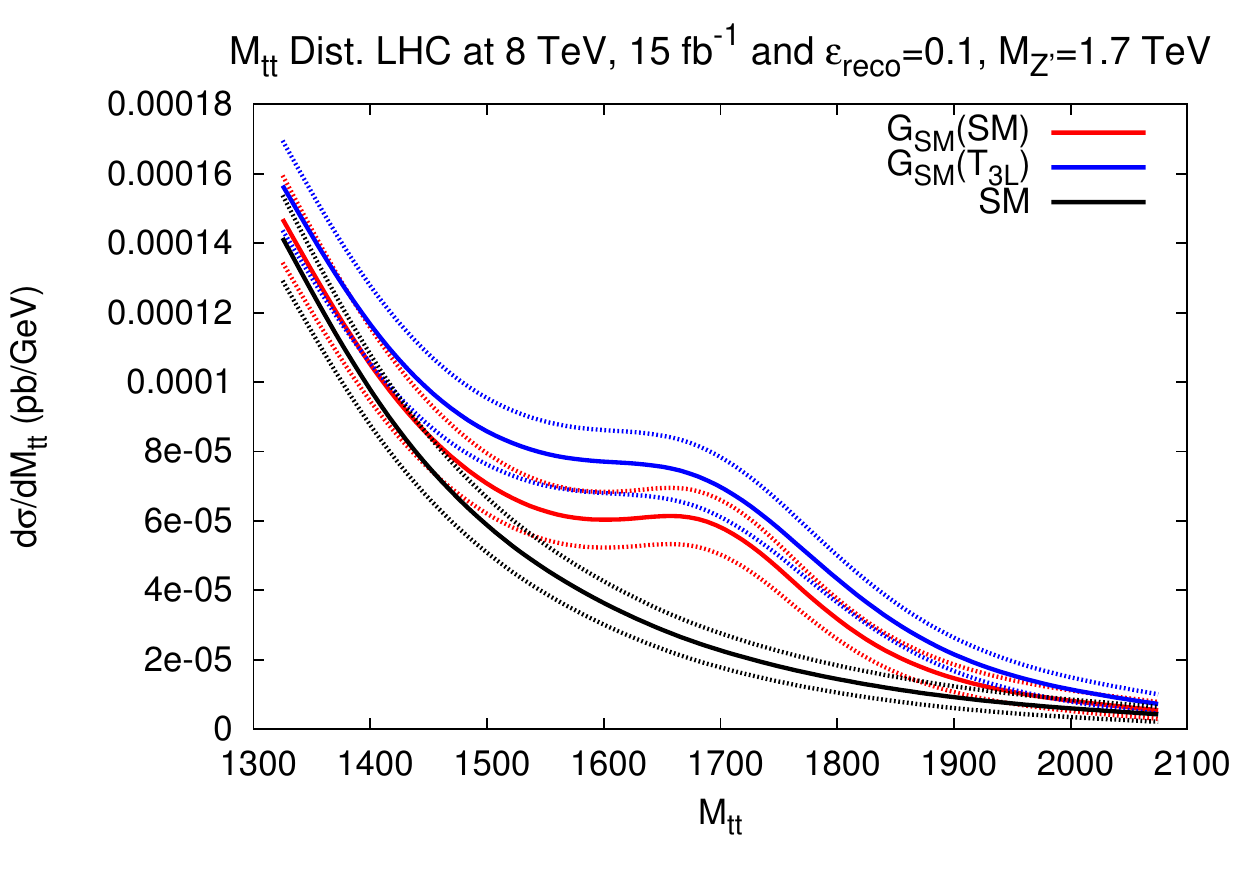}
\caption{Invariant mass distributions for generalised models with $M_{Z^{\prime}}$=1.7 TeV for the LHC at 14(8) TeV and 100(15) fb$^{-1}$ of integrated luminosity.}\label{fig:LHC_GXX_Mtt}
\end{figure}

\begin{figure}[h!]	
	\centering
		\includegraphics[width=0.32\linewidth]{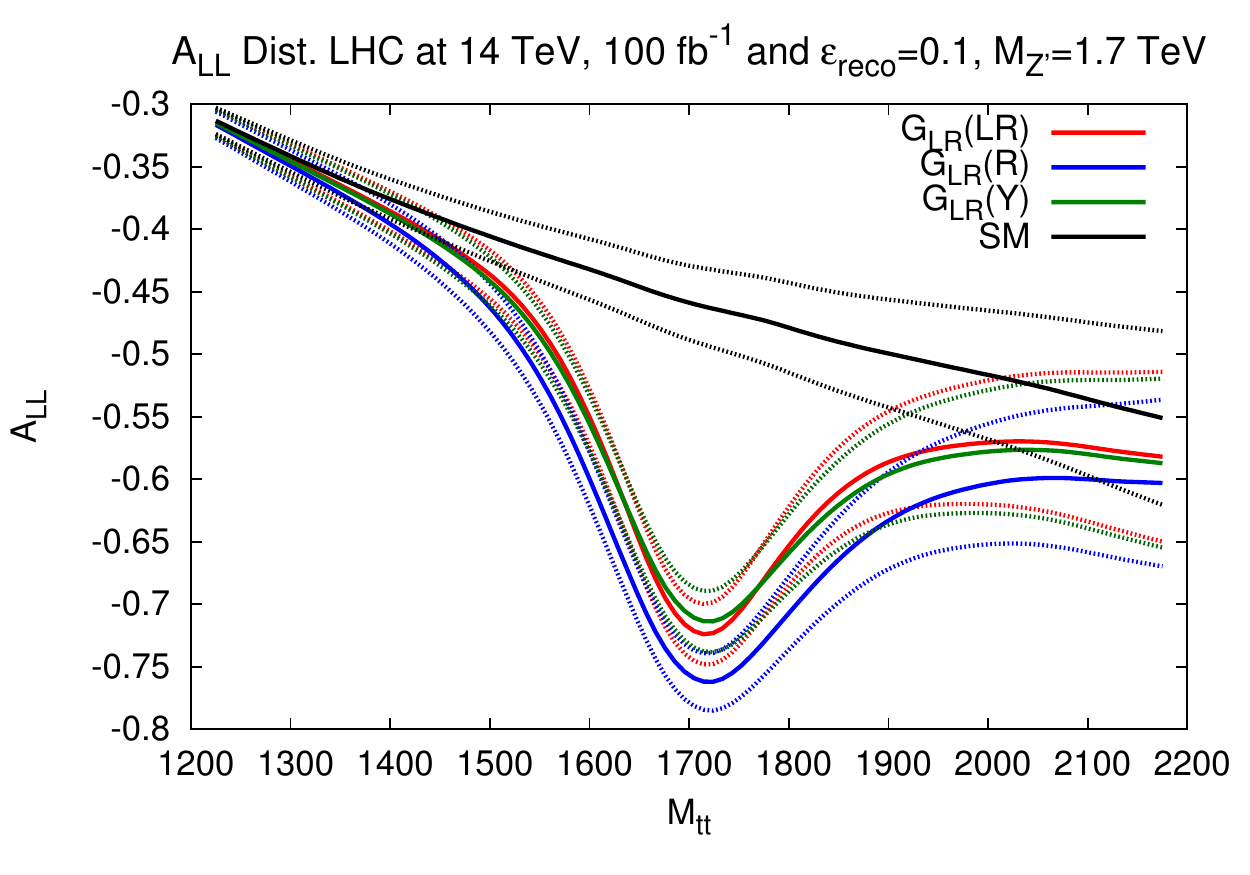}
		\includegraphics[width=0.32\linewidth]{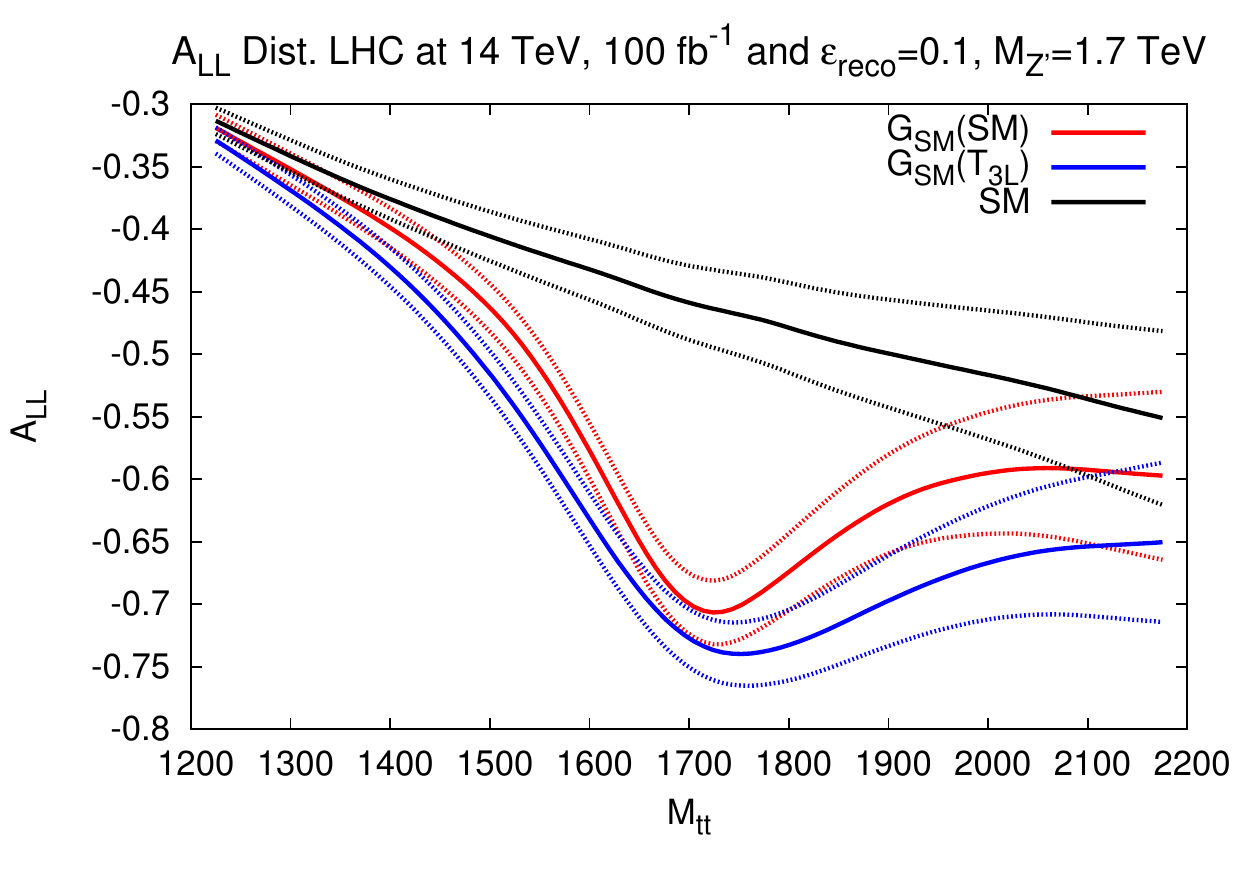}
		\includegraphics[width=0.32\linewidth]{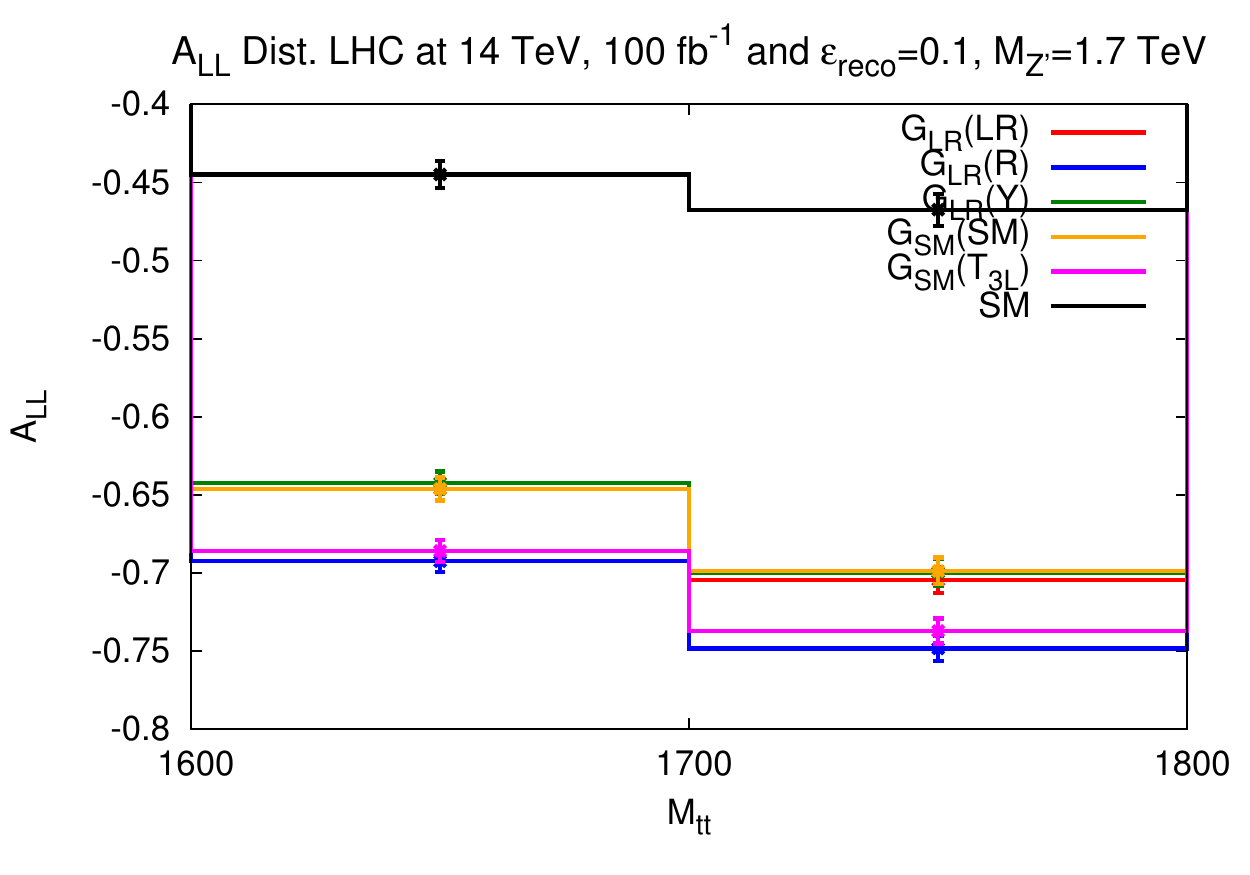}\\
		\includegraphics[width=0.32\linewidth]{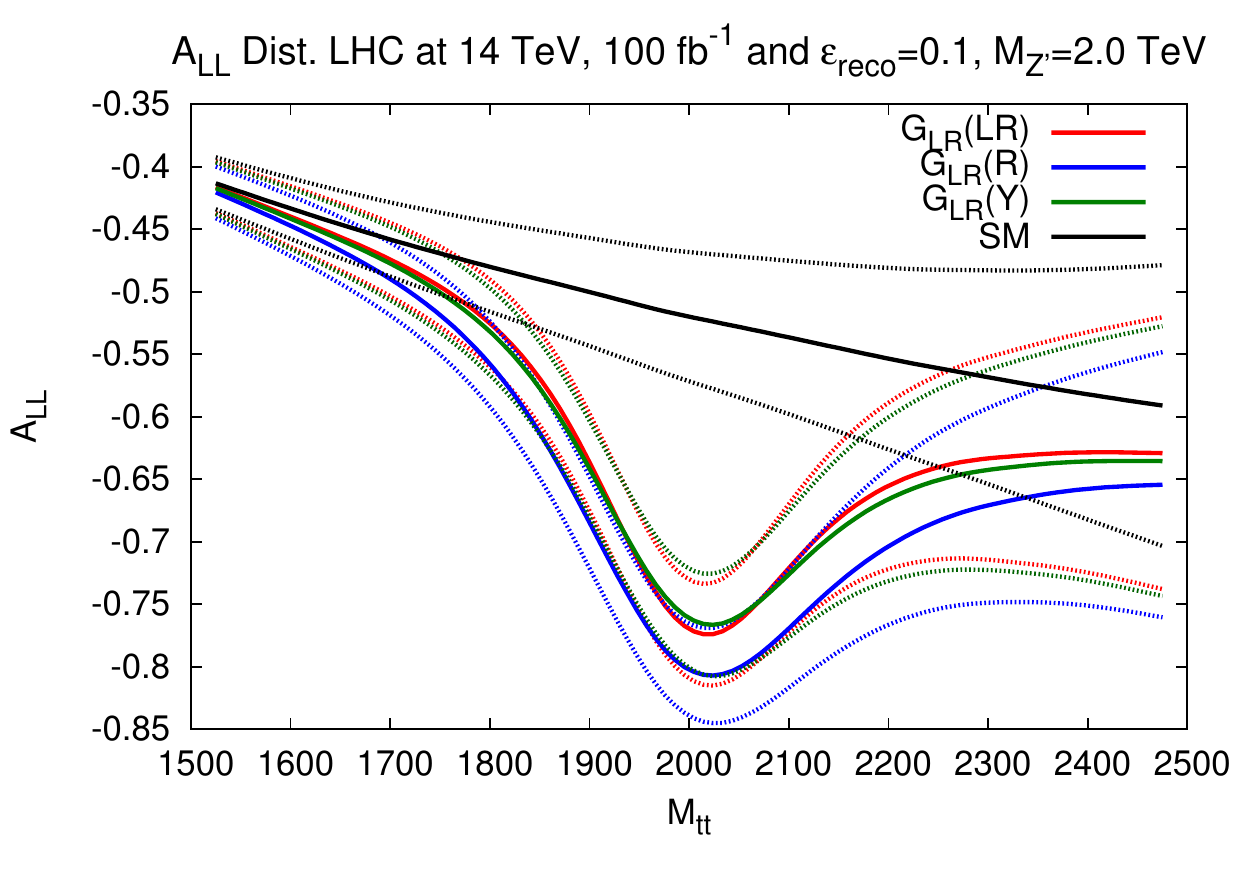}
		\includegraphics[width=0.32\linewidth]{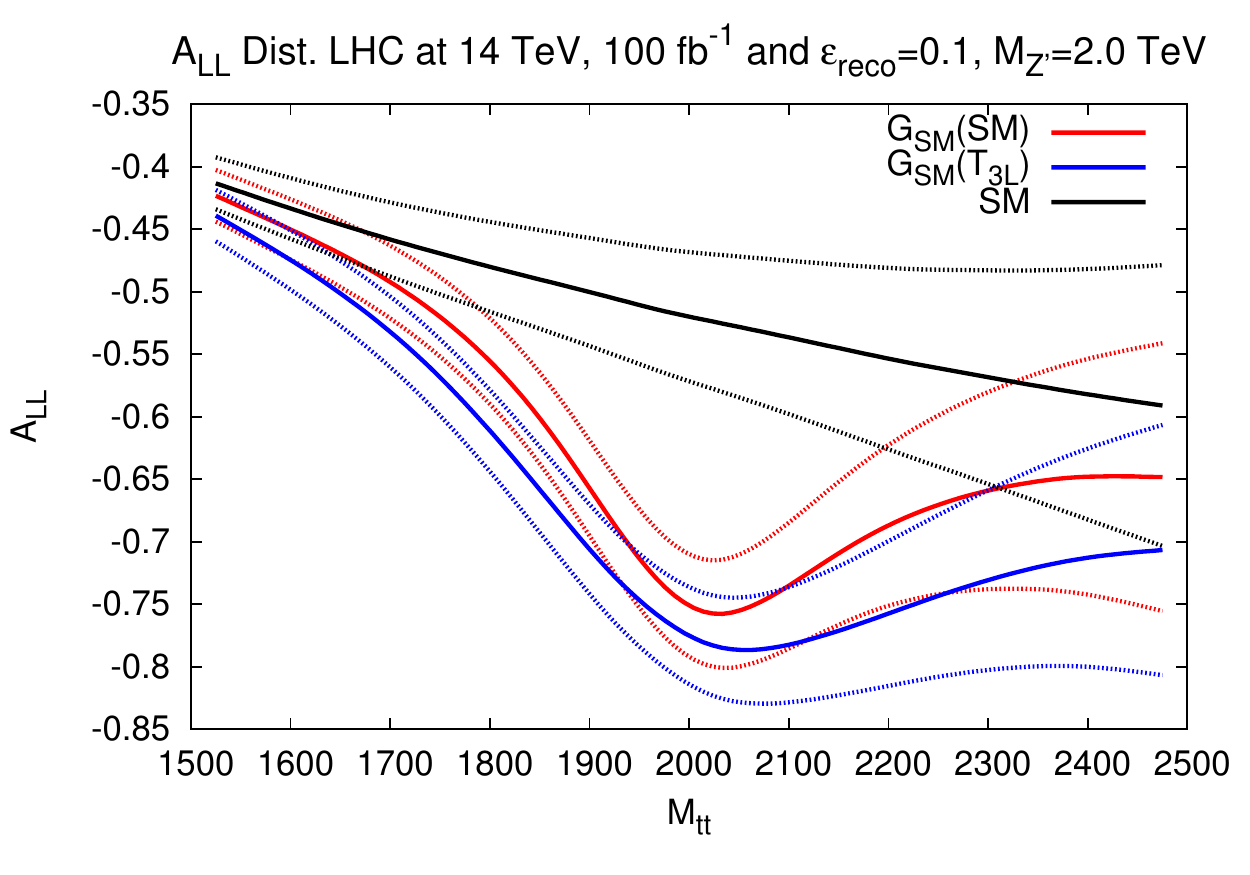}
		\includegraphics[width=0.32\linewidth]{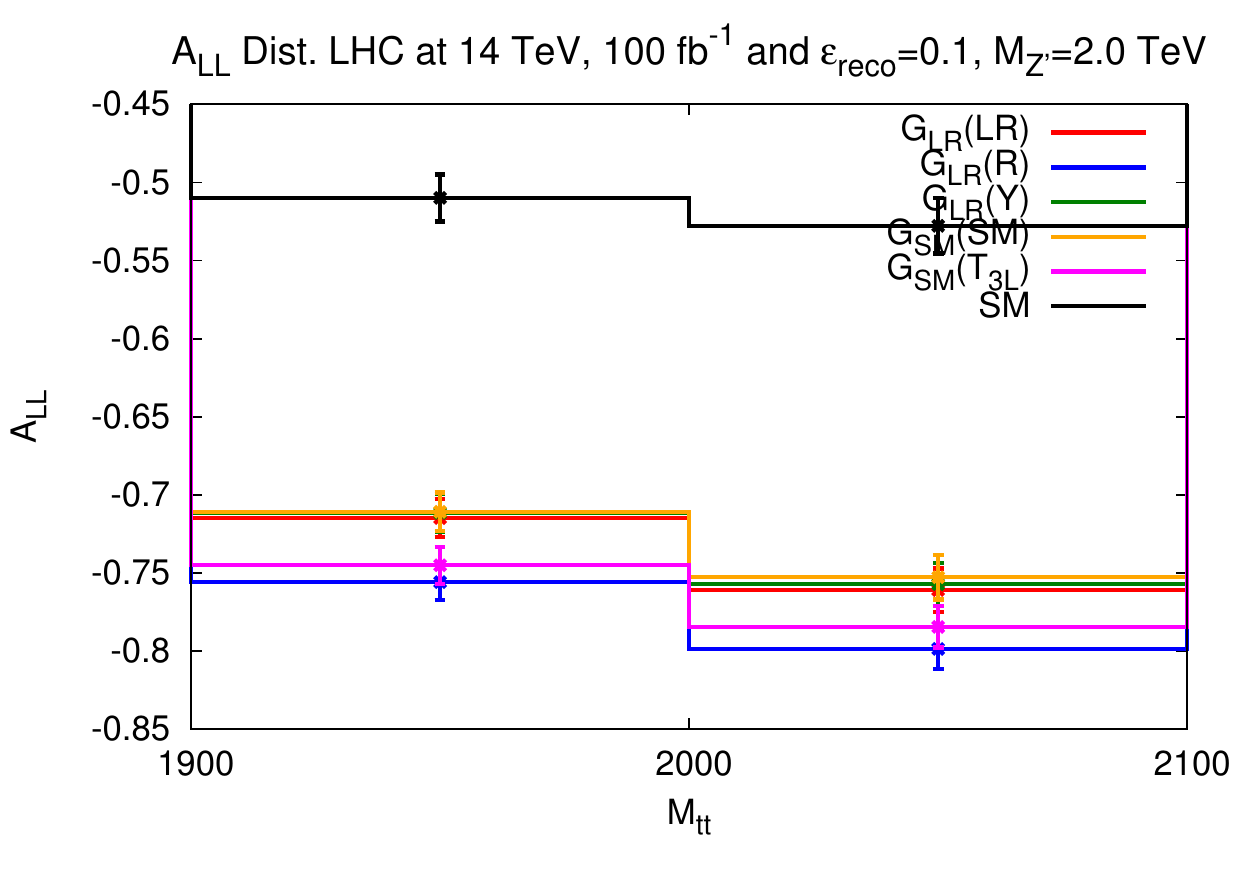}
\caption{$A_{LL}$ distributions binned in $M_{t\overline{t}}$ for generalised models with $M_{Z^{\prime}}$=1.7 (\emph{upper}) and 2 (\emph{lower}) TeV
for the LHC at 14 TeV assuming 100 fb$^{-1}$ of integrated luminosity. Rightmost plots show the distribution in two 100 GeV bins either side of the 
$Z^{\prime}$ peak. Dotted lines and error bars represent statistical uncertainty calculated as described in the text.}\label{fig:LHC14_GXX_ALL}
\end{figure}

\begin{figure}[h!]	
	\centering
		\includegraphics[width=0.32\linewidth]{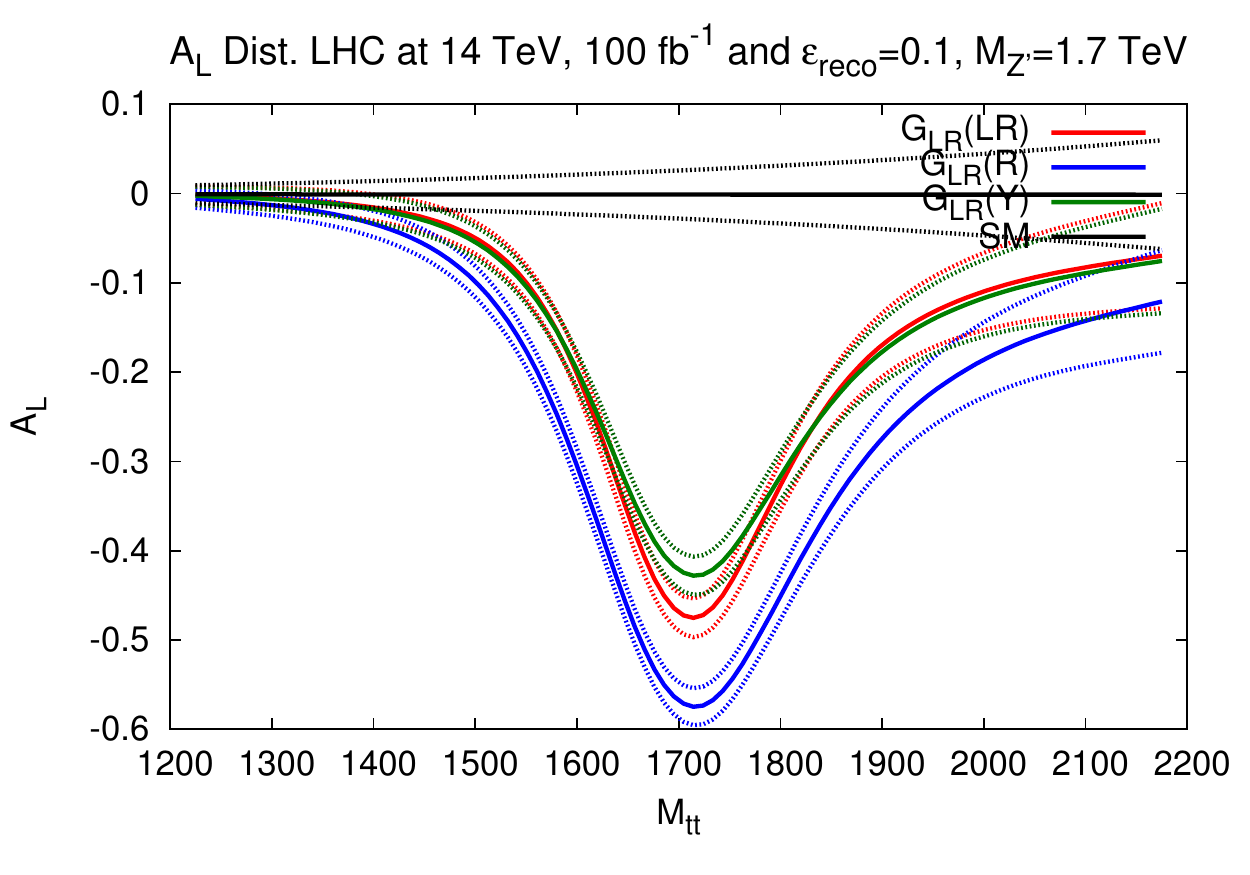}
		\includegraphics[width=0.32\linewidth]{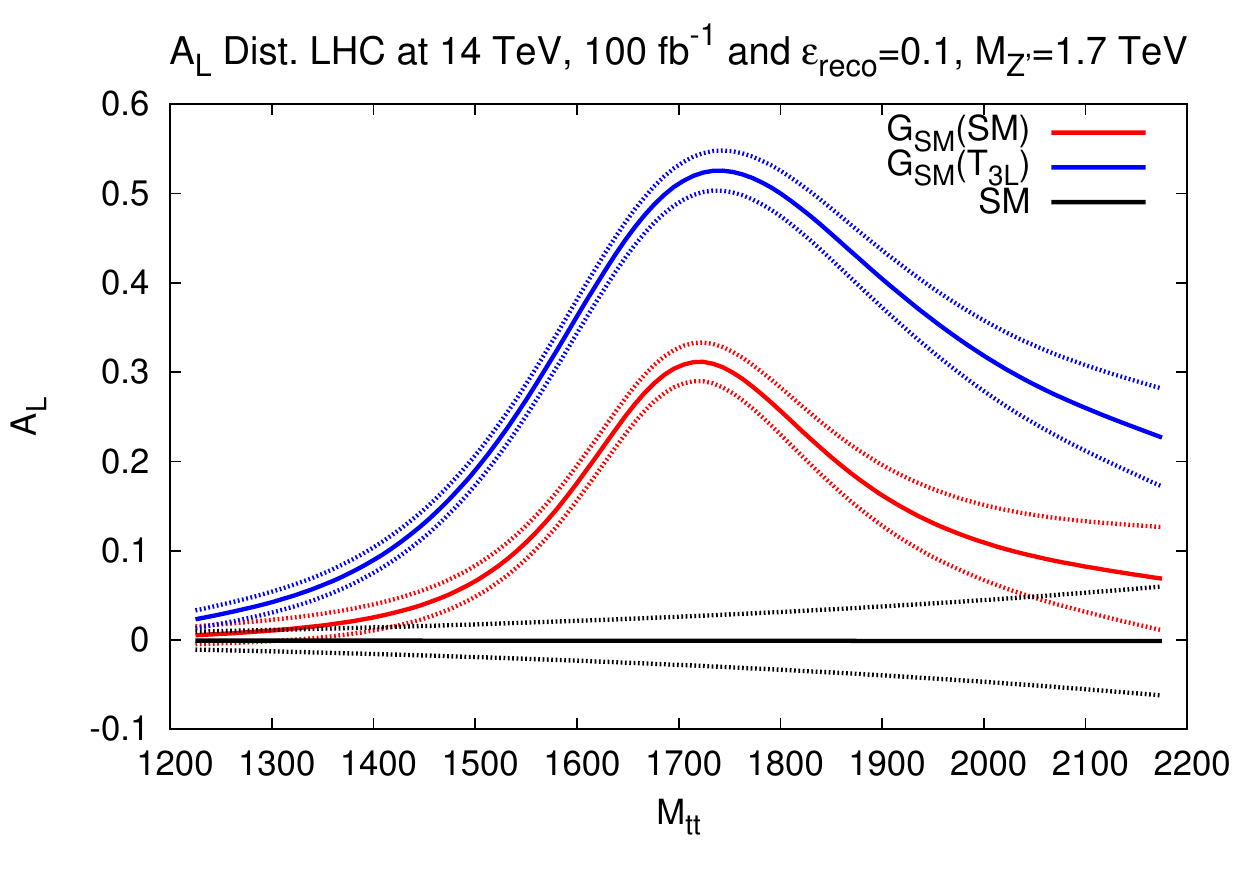}
		\includegraphics[width=0.32\linewidth]{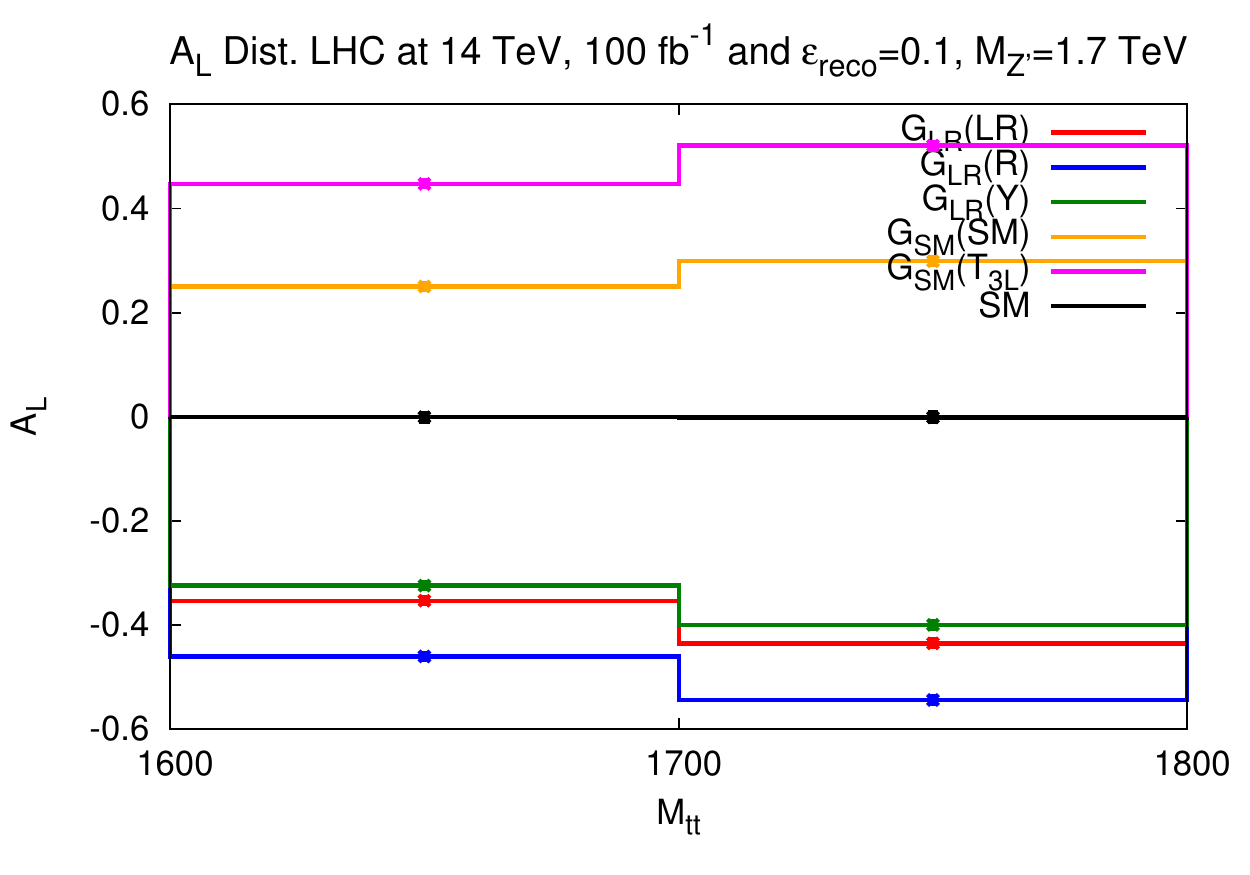}\\
		\includegraphics[width=0.32\linewidth]{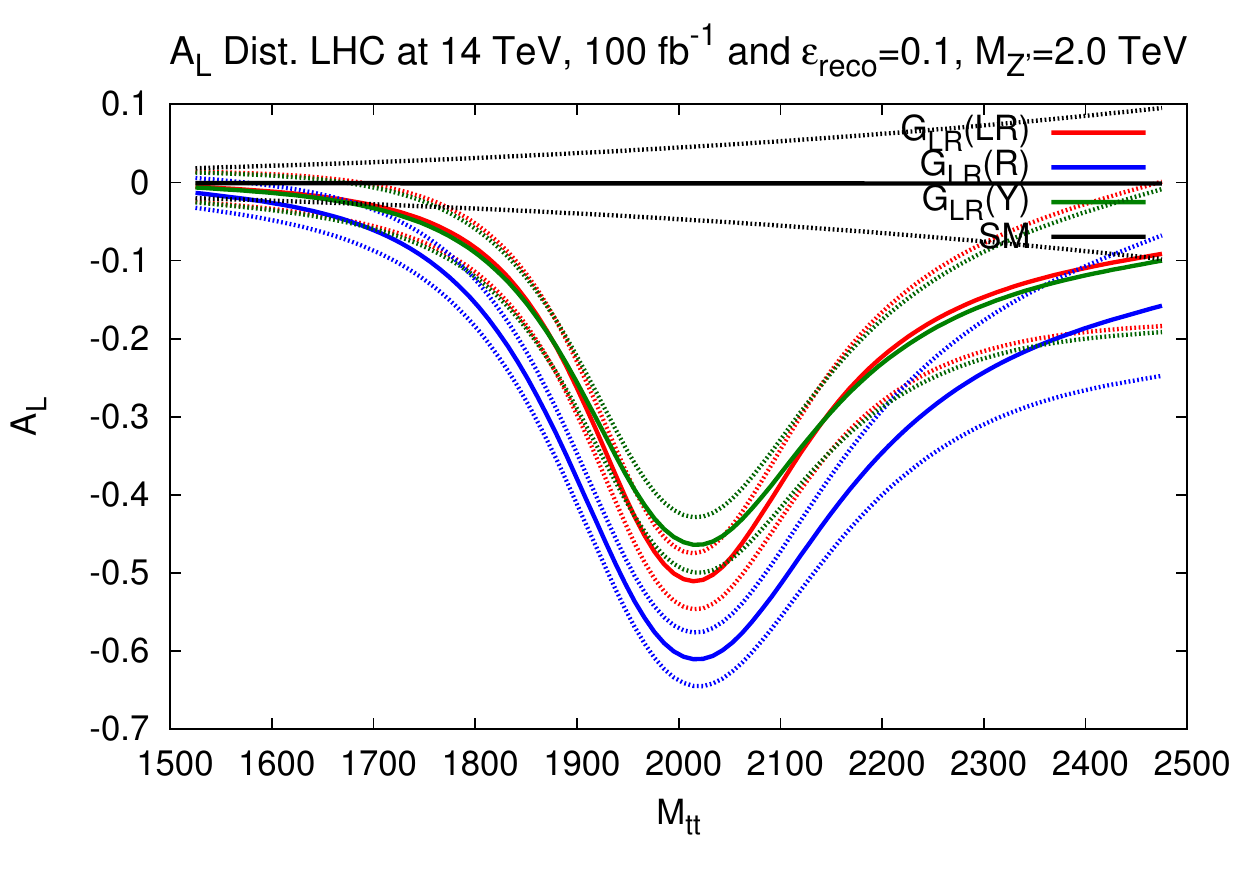}
		\includegraphics[width=0.32\linewidth]{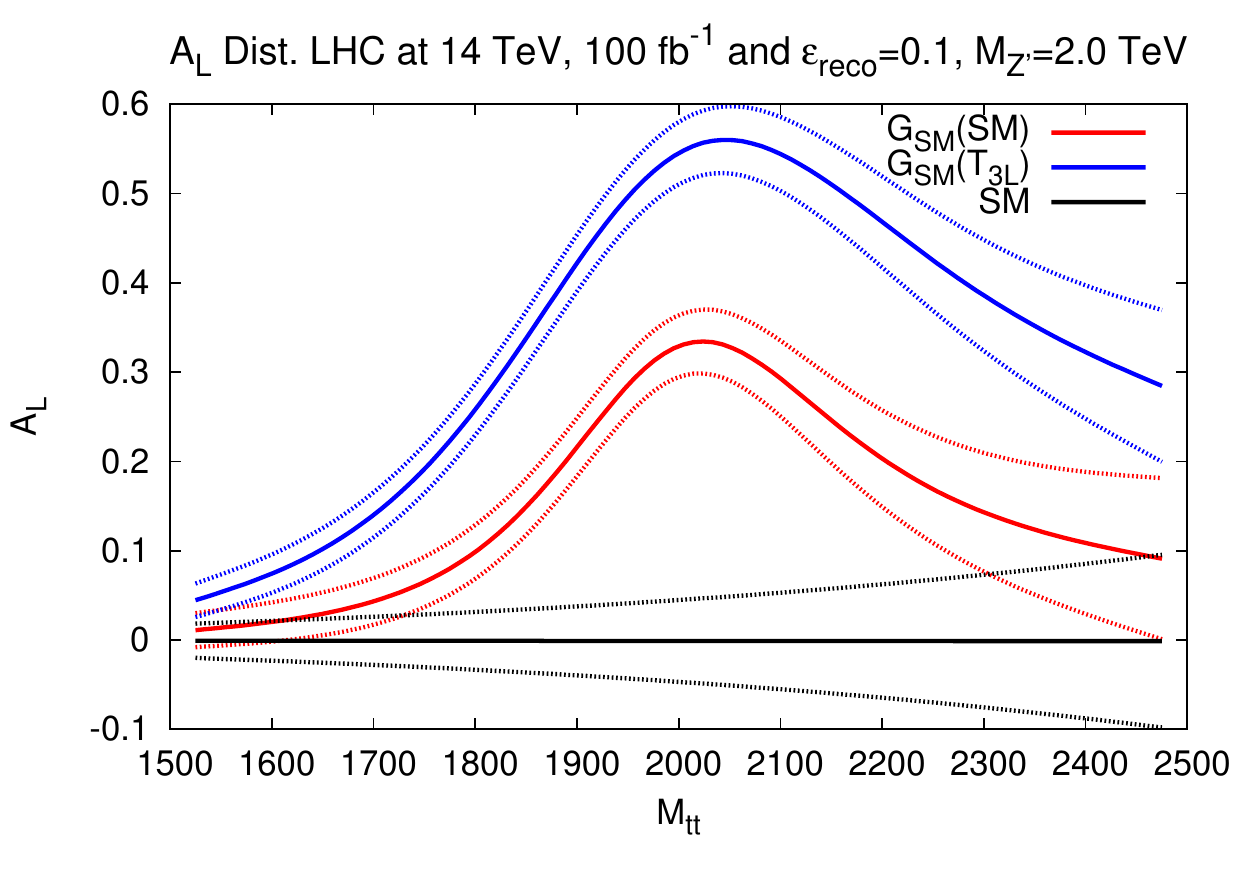}
		\includegraphics[width=0.32\linewidth]{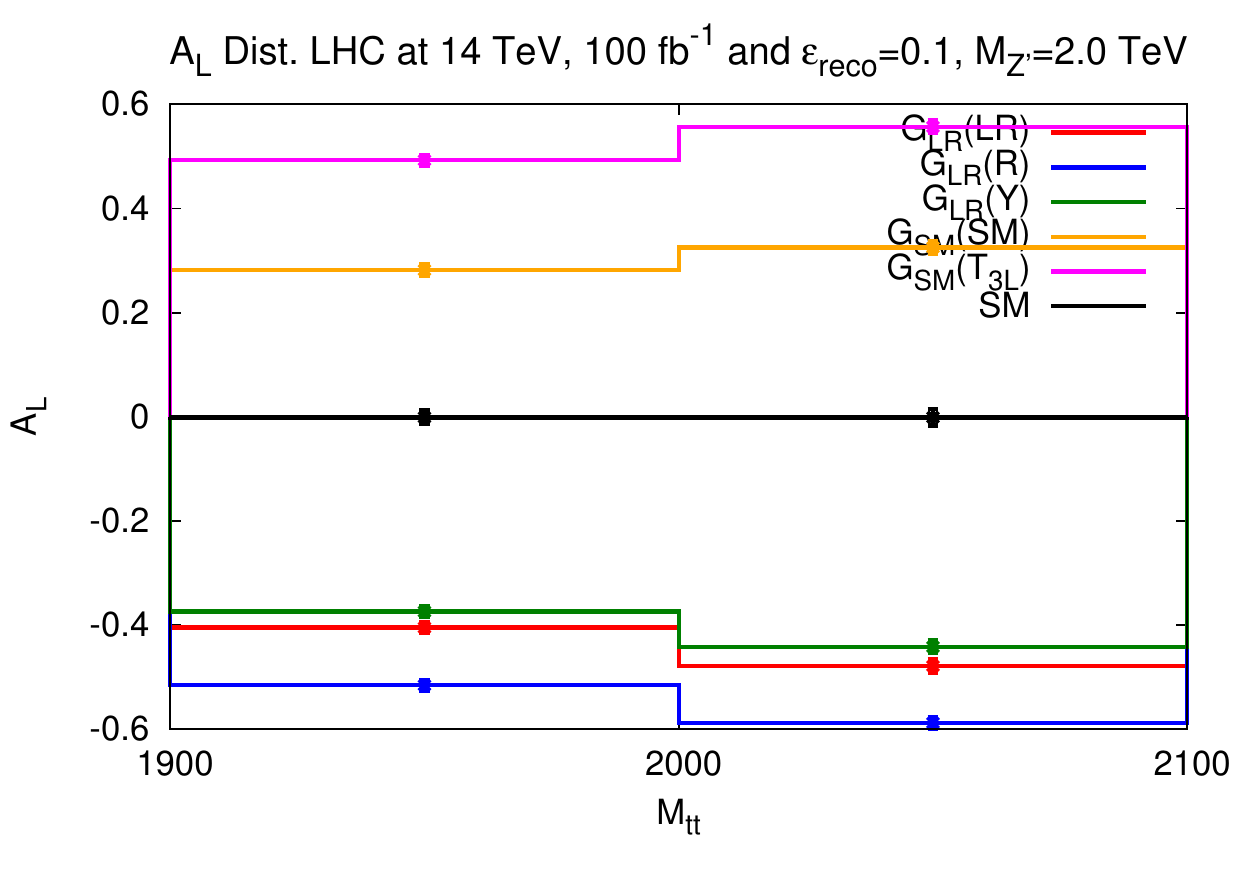}
\caption{$A_{L}$ binned in $M_{t\overline{t}}$ for generalised models with $M_{Z^{\prime}}$=1.7 (\emph{upper}) and 2 (\emph{lower}) TeV
for the LHC at 14 TeV assuming 100 fb$^{-1}$ of integrated luminosity. Rightmost plots show the distribution in two 100 GeV bins either side of the 
$Z^{\prime}$ peak. Dotted lines and error bars represent statistical uncertainty calculated as described in the text.}\label{fig:LHC14_GXX_AL}
\end{figure}

In contrast to the $E_{6}$-type models, the generalised models have non-zero vector and axial couplings to all quarks.
Therefore, all of the asymmetry observables can be generated at tree-level on peak. Combined with their consequently
higher cross sections, shown 
in the invariant mass distributions of Figure~\ref{fig:LHC_GXX_Mtt}, this set of models has clear signatures at the LHC even at 8 TeV. 
Figures~\ref{fig:LHC14_GXX_ALL} and~\ref{fig:LHC14_GXX_AL} and Tables~\ref{tab:LHC_GXX_ALL} and~\ref{tab:LHC_GXX_AL} profile the spin asymmetry variables $A_{LL}$ and 
$A_{L}$, showing large deviations from the $SM$ case. As already noted, the difference in sign in the $A_{L}$ contributions of the 
$G_{LR}$ and $G_{SM}$ models allows for the best distinguishing power over all the models investigated. This is particularly important for the specific case of the $G_{LR}(LR/Y)$ and $G_{SM}(SM)$ models that do not appear distinguishable in the invariant mass distributions nor in the other variables, but do so in the $A_{L}$ two-bin plots.

\begin{figure}[h!]	
	\centering
		\includegraphics[width=0.32\linewidth]{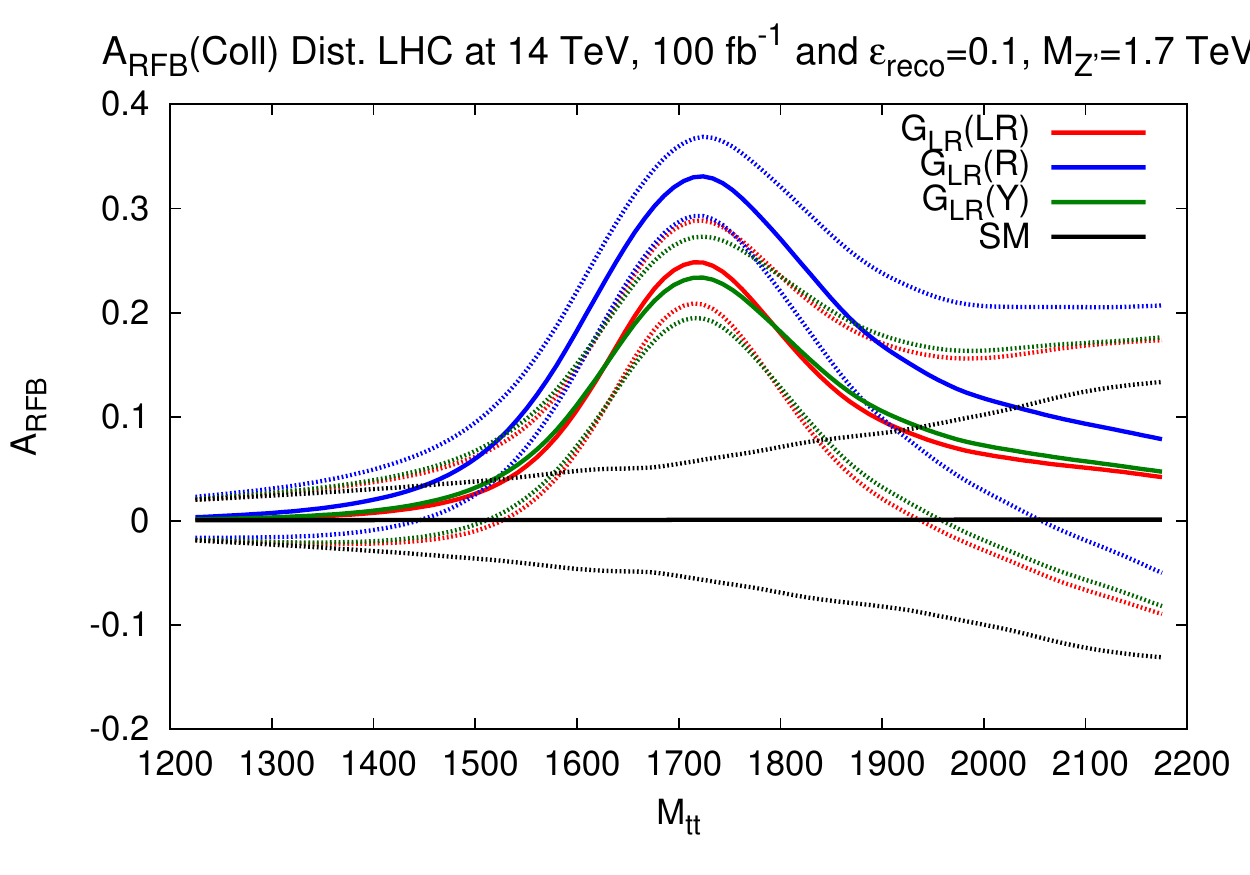}
		\includegraphics[width=0.32\linewidth]{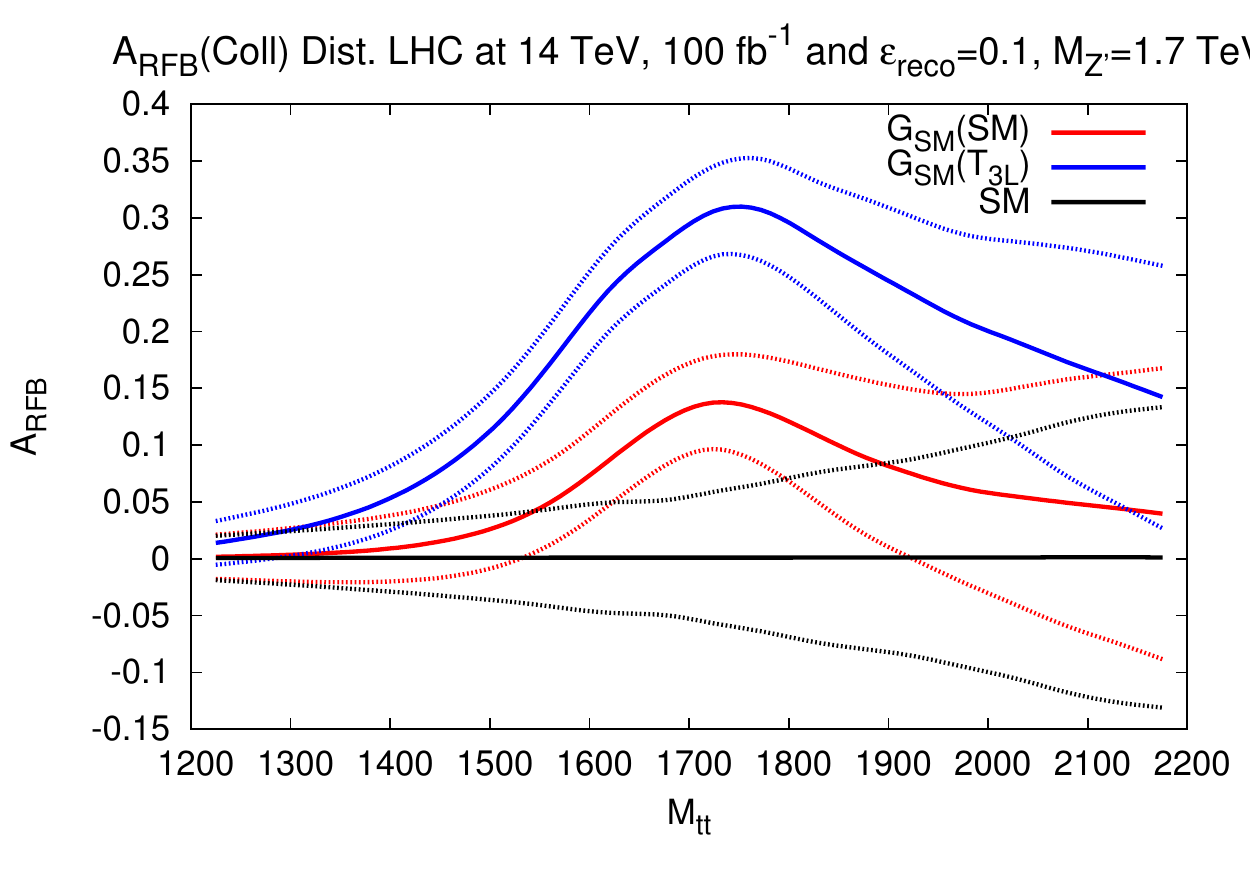}
		\includegraphics[width=0.32\linewidth]{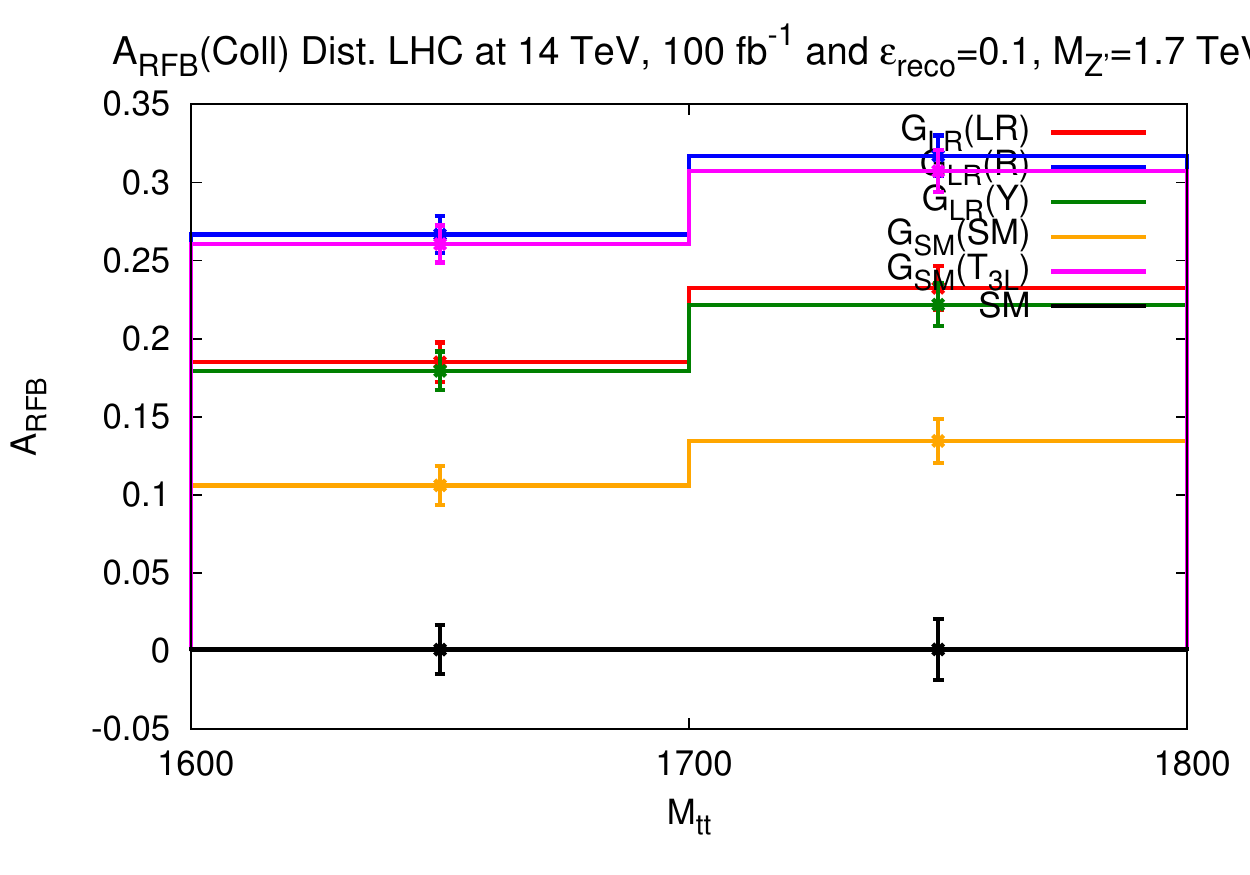}\\
		\includegraphics[width=0.32\linewidth]{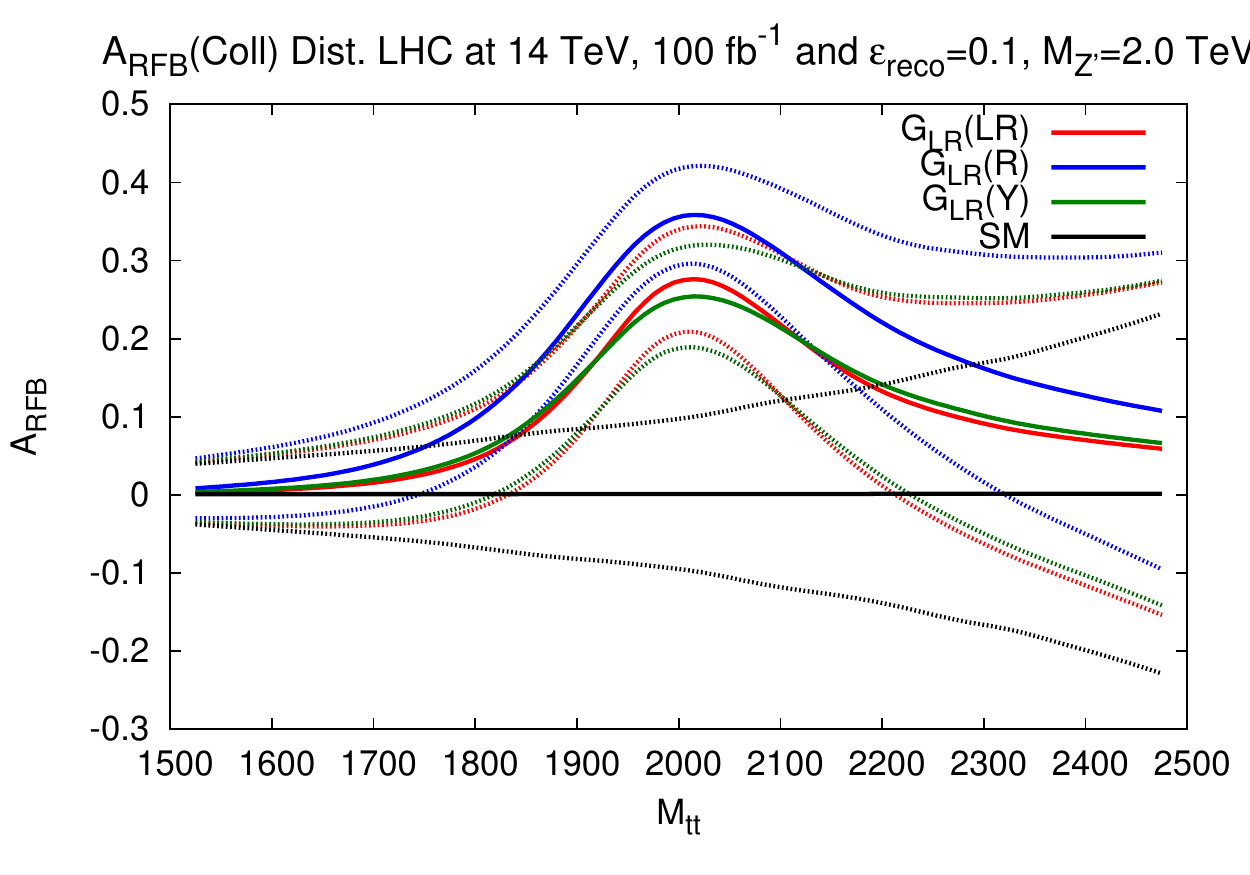}
		\includegraphics[width=0.32\linewidth]{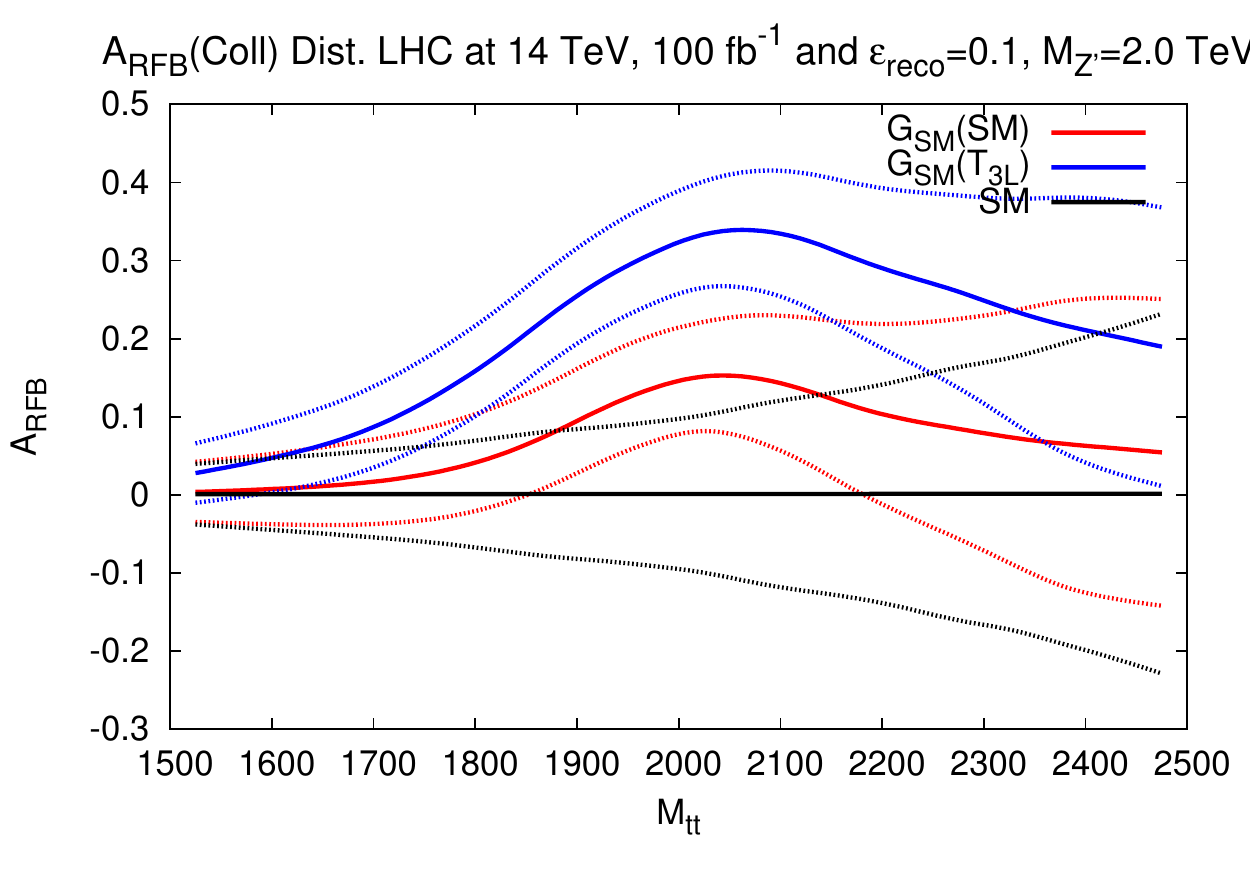}
		\includegraphics[width=0.32\linewidth]{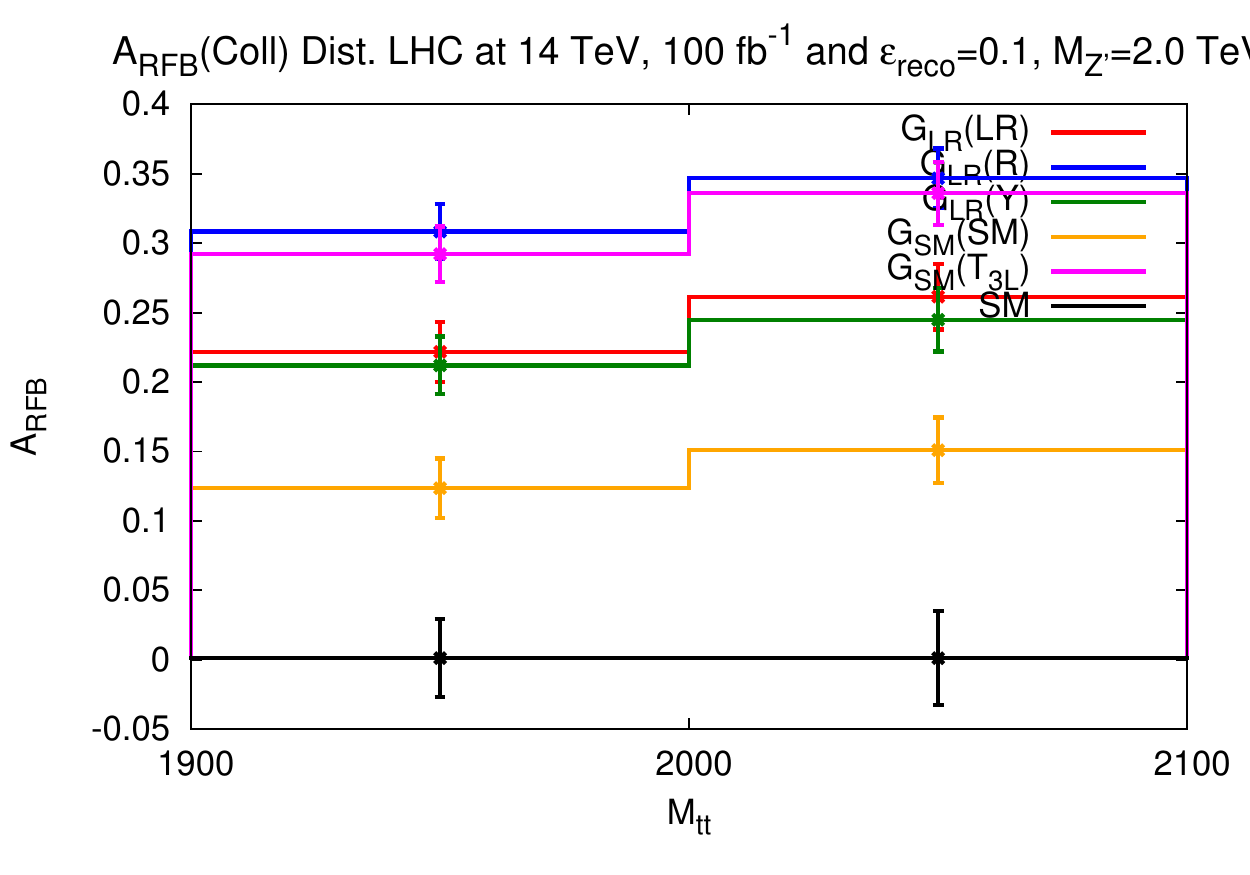}
\caption{$A_{RFB}$ binned in $M_{t\overline{t}}$ for generalised models with $M_{Z^{\prime}}$=1.7 (\emph{upper}) and 2 (\emph{lower}) TeV
for the LHC at 14 TeV assuming 100 fb$^{-1}$ of integrated luminosity. Rightmost plots show the distribution in two 100 GeV bins either side of the 
$Z^{\prime}$ peak. Dotted lines and error bars represent statistical uncertainty calculated as described in the text.}\label{fig:LHC14_GXX_ARFB}
\end{figure}

The spatial asymmetry variables are also clearly visible in these models and since they all are very similar kinematically,
we show plots for the variable with the best discrimination power\footnote{This is expected from the fact that the variable incorporates a kinematical cut to enhance the $q\bar{q}$ contribution as discussed in section~\ref{subsec:chargedef}} only (as we will see in the next section), i.e., $A_{RFB}$, in Figure~\ref{fig:LHC14_GXX_ARFB} in differential form while its integrated values are found in Table~\ref{tab:LHC_GXX_ARFB}.

\subsection{Significance and luminosity analysis}
Tables~\ref{tab:signif_LHC14_ALL} to~\ref{tab:signif_LHC14_ARFB} summarise the significance measures between various models as defined in eq.~(\ref{eqn:signif}) for $A_{LL}$, $A_{L}$ and $A_{RFB}$, for the entries in Tables~\ref{tab:LHC_E_6_ALL} to \ref{tab:LHC_GXX_ARFB}, respectively.

Generally speaking, $A_L$ provides the best overall discrimination power, when the variable is non-vanishing at the $Z'$ peak. Beside preserving the relative sign between the top quark chiral couplings, allowing to distinguish the $G_{SM}$ from the $G_{LR}$ class, it also has the highest significance when comparing the $Z'$ models in these classes with the SM expectation and among themselves. For the $E_6$-type cases, where $A_L$ is too small to be measured, the other spin variable $A_{LL}$ comes into play. We observe that its significance is always bigger than the spatial asymmetries, $A_{RFB}$ being the biggest amongst the latter. As mentioned, this variable can be used to distinguish all the presented models from the SM background. Regarding the disentanglement among models, we observe that the narrower mass window and lower $Z^{\prime}$ mass always perform better, although noting that 1.7 TeV is bordering on being ruled out already in other searches. Models with very large width, such as the $G_{SM}$ ones, deliver a bigger significance in the larger invariant mass cut. Finally, The significance of these asymmetry measures is finally evaluated for the case of both $E_6$ and generalised models
in Tables~\ref{tab:signif_LHC14_ALL}--\ref{tab:signif_LHC14_ARFB}, for $A_{LL}$, $A_L$ and $A_{RFB}$.

From all these tables it is clear that some models can be distinguished in early stages of the LHC at $\sqrt{s}=14$ TeV, i.e., with less than 100 fb$^{-1}$. At the same time, the question of what is the required luminosity to discriminate among models also quantify how powerful a variable is. Tables~\ref{tab:Lumi_LHC14_ALL-AL} and~\ref{tab:Lumi_LHC14_spatial} address the distinguishability of the various models using the spin and spatial asymmetries with increasing integrated luminosity. These reinforce the fact that the models can generally be separated using these observables for reasonable integrated luminosities when the sizes (and signs) of the relevant couplings differ enough. The spin asymmetries provide the best distinctions and $A_{RFB}$ performs the best among the spatial asymmetries. 

Although certain models remain unresolvable even with full luminosity, $300$ fb$^{-1}$, $\mathcal{O}(1)$ fb$^{-1}$ of integrated luminosity is already enough to disentangle the generalised models using $A_L$. With $A_{LL}$, $E_6$-type models start being distinguishable with $\mathcal{O}(10)$ fb$^{-1}$. $\mathcal{O}(50)$ fb$^{-1}$ is required for full discrimination with $A_{LL}$, as well as to have confirmations for the generalised models with the spatial asymmetries, among which $A_{RFB}$ outperforms all the others requiring less integrated luminosity.

\section{Conclusions \label{sec:summary}}

$Z'$'s appears in many BSM scenarios and are under intense scrutiny presently at the LHC owning to a clear leptonic signature they offer
through the DY process. However, in order to profile this new object
(i.e., measuring not only its mass and width but also its couplings
to all matter, its spin, its $CP$ quantum numbers, etc.), a combination between the DY results and those obtained in other 
channels may be necessary. Channels where asymmetries in the event
distribution can be defined are particularly powerful, as such observables are very sensitive to the chiral structure of
the $Z'$-fermion-antifermion vertices involved and to the spin nature
of the new particle. Hence, amongst these, the $t\bar t$ decay channel of a $Z'$ boson is particularly relevant and accessible at the LHC, albeit less clean and efficient that the DY ones.

We have therefore presented a phenomenological study of classes of $Z^{\prime}$ models 
in both spin and spatial asymmetries of $t\bar{t}$ production. A selection of observables has been 
defined and profiled as a function of the $t\bar t$ invariant mass showing that there is much scope to observe 
deviations from the SM and even distinguish between various models, particularly for spin asymmetries, using a 
narrow invariant mass range around the $Z^\prime$ peak. Further, we quantified distinguishability between 
models and considered the significance of such differences with respect to the integrated luminosity.

It is worth noting that, as stated in Section~\ref{sect:intro}, the classes of models studied are a set 
of benchmarks put forward for experimentalists to set bounds on $Z^{\prime}$ masses which are best 
probed in the di-lepton channels. Other models featuring heavy neutral gauge bosons would be even better 
suited to the $t\bar{t}$ channel, such as leptophobic/top-phillic $Z^{\prime}$s occurring in composite/multi-site and 
extra-dimensional models \cite{topfriendly,prep}. The profiling techniques discussed in this study would be increasingly more 
applicable in these top-friendly scenarios. Furthermore, it has been assumed that all other exotic matter 
states decoupled for simplicity while they may have non-negligible effects on widths and branching ratios 
that should be considered when moving away from model independent methods \cite{prep}.

\clearpage
	\begin{table}[h!]
		\centering
		\begin{tabular}{|c|cc|cc|}
		\hline
		$A_{LL}(\times 10)$&$\sqrt{s}=14$ TeV&$\mathcal{L}_{int}=100$ fb$^{-1}$&$\sqrt{s}=8$ TeV&$\mathcal{L}_{int}=15$ fb$^{-1}$\\
		\hline
		\hline
		$M_{Z^{\prime}}=1.7$ TeV&$\Delta M_{t\bar t}<0.5$ TeV&$\Delta M_{t\bar t}<0.1$ TeV&$\Delta M_{t\bar t}<0.5$ TeV&$\Delta M_{t\bar t}<0.1$ TeV\\
		\hline
			$SM$&$-3.79\pm0.05$&$-4.54\pm0.07$&$-4.75\pm0.39$&$-5.65\pm0.61$\\
			$E_{6}(\chi)$&$-3.88\pm0.05$&$-5.07\pm0.06$&$-4.85\pm0.39$&$-6.35\pm0.58$\\
			$E_{6}(\eta)$&$-4.17\pm0.05$&$-6.42\pm0.06$&$-5.22\pm0.38$&$-7.85\pm0.48$\\
			$E_{6}(\psi)$&$-4.01\pm0.05$&$-5.79\pm0.06$&$-5.02\pm0.33$&$-7.22\pm0.52$\\
			$E_{6}(N)$&$-3.90\pm0.05$&$-5.21\pm0.06$&$-4.88\pm0.39$&$-6.54\pm0.57$\\
			$E_{6}(S)$&$-3.80\pm0.05$&$-4.62\pm0.07$&$-4.76\pm0.39$&$-5.76\pm0.61$\\
			$G_{LR}(B-L)$&$-3.88\pm0.05$&$-5.02\pm0.06$&$-4.86\pm0.39$&$-6.31\pm0.57$\\
		\hline
		\hline
		$M_{Z^{\prime}}=2.0$ TeV&$\Delta M_{t\bar t}<0.5$ TeV&$\Delta M_{t\bar t}<0.1$ TeV&$\Delta M_{t\bar t}<0.5$ TeV&$\Delta M_{t\bar t}<0.1$ TeV\\
		\hline
			$SM$&$-4.66\pm0.09$&$-5.17\pm0.11$&$-5.68\pm0.84$&$-6.32\pm1.23$\\
			$E_{6}(\chi)$&$-4.77\pm0.09$&$-5.76\pm0.11$&$-5.81\pm0.83$&$-7.03\pm1.14$\\
			$E_{6}(\eta)$&$-5.13\pm0.09$&$-7.15\pm0.10$&$-6.26\pm0.80$&$-8.44\pm0.89$\\
			$E_{6}(\psi)$&$-4.94\pm0.09$&$-6.54\pm0.10$&$-6.02\pm0.82$&$-7.90\pm1.00$\\
			$E_{6}(N)$&$-4.79\pm0.09$&$-5.92\pm0.11$&$-5.84\pm0.83$&$-7.23\pm1.11$\\
			$E_{6}(S)$&$-4.67\pm0.09$&$-5.27\pm0.11$&$-5.70\pm0.84$&$-6.43\pm1.22$\\
			$G_{LR}(B-L)$&$-4.77\pm0.09$&$-5.70\pm0.11$&$-5.82\pm0.83$&$-7.00\pm1.13$\\
		\hline
		\end{tabular}
	\caption{Summary of integrated $A_{LL}$ values around the $Z^{\prime}$ peak for $E_{6}$-type models with $M_{Z^{\prime}}$=1.7 
	and 2 TeV at the LHC at 14 and 8 TeV assuming 100 and 15 fb$^{-1}$ of integrated luminosity respectively.}\label{tab:LHC_E_6_ALL}
	\end{table}

\newpage

\begin{table}[h!]
	\centering
	\begin{tabular}{|c|cc|cc|}
	\hline
	$A_{LL}(\times 10)$&$\sqrt{s}=14$ TeV&$\mathcal{L}_{int}=100$ fb$^{-1}$&$\sqrt{s}=8$ TeV&$\mathcal{L}_{int}=15$ fb$^{-1}$\\
	\hline
	\hline
	$M_{Z^{\prime}}=1.7$ TeV&$\Delta M_{t\bar t}<0.5$ TeV&$\Delta M_{t\bar t}<0.1$ TeV&$\Delta M_{t\bar t}<0.5$ TeV&$\Delta M_{t\bar t}<0.1$ TeV\\
	\hline
	$SM$&$-3.79\pm0.05$&$-4.54\pm0.07$&$-4.75\pm0.39$&$-5.65\pm0.61$\\
	$G_{LR}(LR)$&$-4.41\pm0.05$&$-6.72\pm0.06$&$-5.48\pm0.37$&$-8.03\pm0.45$\\
	$G_{LR}(R)$&$-4.70\pm0.05$&$-7.18\pm0.05$&$-5.83\pm0.36$&$-8.38\pm0.41$\\
	$G_{LR}(Y)$&$-4.43\pm0.05$&$-6.68\pm0.05$&$-5.55\pm0.37$&$-8.02\pm0.44$\\
	$G_{SM}(SM)$&$-4.52\pm0.05$&$-6.69\pm0.06$&$-5.64\pm0.37$&$-8.04\pm0.45$\\
	$G_{SM}(T_{3L})$&$-4.94\pm0.04$&$-7.09\pm0.05$&$-6.12\pm0.35$&$-8.31\pm0.41$\\
	\hline
	\hline
	$M_{Z^{\prime}}=2.0$ TeV&$\Delta M_{t\bar t}<0.5$ TeV&$\Delta M_{t\bar t}<0.1$ TeV&$\Delta M_{t\bar t}<0.5$ TeV&$\Delta M_{t\bar t}<0.1$ TeV\\
	\hline
	$SM$&$-4.66\pm0.09$&$-5.17\pm0.11$&$-5.68\pm0.84$&$-6.32\pm1.23$\\
	$G_{LR}(LR)$&$-5.41\pm0.08$&$-7.36\pm0.09$&$-6.53\pm0.78$&$-8.51\pm0.85$\\
	$G_{LR}(R)$&$-5.74\pm0.08$&$-7.75\pm0.09$&$-6.90\pm0.75$&$-8.79\pm0.76$\\\
	$G_{LR}(Y)$&$-5.44\pm0.08$&$-7.32\pm0.09$&$-6.62\pm0.77$&$-8.53\pm0.82$\\
	$G_{SM}(SM)$&$-5.53\pm0.08$&$-7.30\pm0.09$&$-6.69\pm0.77$&$-8.51\pm0.86$\\
	$G_{SM}(T_{3L})$&$-5.99\pm0.08$&$-7.63\pm0.09$&$-7.16\pm0.72$&$-8.72\pm0.78$\\
	\hline
	\end{tabular}
\caption{Summary of integrated $A_{LL}$ values around the $Z^{\prime}$ peak for the generalised models with $M_{Z^{\prime}}$=1.7 
and 2 TeV at the LHC at 14 and 8 TeV assuming 100 and 15 fb$^{-1}$ of integrated luminosity respectively.}\label{tab:LHC_GXX_ALL}
\end{table}

\begin{table}[h!]
	\centering
	\begin{tabular}{|c|cc|cc|}
	\hline
	$A_{L}(\times 10)$&$\sqrt{s}=14$ TeV&$\mathcal{L}_{int}=100$ fb$^{-1}$&$\sqrt{s}=8$ TeV&$\mathcal{L}_{int}=15$ fb$^{-1}$\\
	\hline
	\hline
	$M_{Z^{\prime}}=1.7$ TeV&$\Delta M_{t\bar t}<0.5$ TeV&$\Delta M_{t\bar t}<0.1$ TeV&$\Delta M_{t\bar t}<0.5$ TeV&$\Delta M_{t\bar t}<0.1$ TeV\\
	\hline
	$SM$&$-0.009\pm0.044$&$-0.010\pm0.059$&$-0.017\pm0.35$&$-0.020\pm0.53$\\
	$G_{LR}(LR)$&$-0.971\pm0.042$&$-3.90\pm0.05$&$-1.37\pm0.33$&$-5.36\pm0.40$\\
	$G_{LR}(R)$&$-1.51\pm0.04$&$-4.98\pm0.05$&$-2.14\pm0.32$&$-6.53\pm0.37$\\
	$G_{LR}(Y)$&$-0.938\pm0.042$&$-3.58\pm0.05$&$-1.40\pm0.33$&$-5.05\pm0.38$\\
	$G_{SM}(SM)$&$0.802\pm0.042$&$2.71\pm0.05$&$1.16\pm0.32$&$3.79\pm0.38$\\
	 $G_{SM}(T_{3L})$&$1.90\pm0.04$&$4.80\pm0.05$&$2.70\pm0.31$&$6.36\pm0.38$\\ 
	\hline
	\hline
	$M_{Z^{\prime}}=2.0$ TeV&$\Delta M_{t\bar t}<0.5$ TeV&$\Delta M_{t\bar t}<0.1$ TeV&$\Delta M_{t\bar t}<0.5$ TeV&$\Delta M_{t\bar t}<0.1$ TeV\\
	\hline
	$SM$&$-0.011\pm0.088$&$-0.012\pm0.10$&$-0.020\pm0.73$&$-0.020\pm1.04$\\
	$G_{LR}(LR)$&$-1.38\pm0.07$&$-4.38\pm0.08$&$-1.91\pm0.66$&$-5.81\pm0.75$\\
	$G_{LR}(R)$&$-2.09\pm0.07$&$-5.49\pm0.08$&$-2.91\pm0.64$&$-6.97\pm0.69$\\
	$G_{LR}(Y)$&$-1.34\pm0.07$&$-4.05\pm0.08$&$-1.99\pm0.65$&$-5.54\pm0.71$\\
	$G_{SM}(SM)$&$1.12\pm0.07$&$3.01\pm0.08$&$1.59\pm0.65$&$4.07\pm0.71$\\
	$G_{SM}(T_{3L})$&$2.55\pm0.07$&$5.21\pm0.08$&$3.53\pm0.62$&$6.74\pm0.71$\\
	\hline
	\end{tabular}
\caption{Summary of integrated $A_{L}$ values around the $Z^{\prime}$ peak for the generalised models with $M_{Z^{\prime}}$=1.7 
and 2 TeV at the LHC at 14 and 8 TeV assuming 100 and 15 fb$^{-1}$ of integrated luminosity respectively.}\label{tab:LHC_GXX_AL}
\end{table}

\begin{table}[h!]
	\centering
	\begin{tabular}{|c|cc|cc|}
	\hline
	$A_{RFB}(\times 10)$&$\sqrt{s}=14$ TeV&$\mathcal{L}_{int}=100$ fb$^{-1}$&$\sqrt{s}=8$ TeV&$\mathcal{L}_{int}=15$ fb$^{-1}$\\
	\hline
	\hline
	$M_{Z^{\prime}}=1.7$ TeV&$\Delta M_{t\bar t}<0.5$ TeV&$\Delta M_{t\bar t}<0.1$ TeV&$\Delta M_{t\bar t}<0.5$ TeV&$\Delta M_{t\bar t}<0.1$ TeV\\
	\hline
	$SM$&$0.008\pm0.089$&$0.010\pm0.122$&$0.018\pm0.819$&$0.02\pm1.35$\\
	$G_{LR}(LR)$&$0.501\pm0.084$&$2.07\pm0.09$&$0.776\pm0.760$&$3.27\pm0.88$\\
	$G_{LR}(R)$&$0.873\pm0.081$&$2.89\pm0.09$&$1.34\pm0.73$&$4.24\pm0.79$\\
	$G_{LR}(Y)$&$0.523\pm0.083$&$1.99\pm0.09$&$0.807\pm0.745$&$2.96\pm0.82$\\
	$G_{SM}(SM)$&$0.337\pm0.083$&$1.19\pm0.09$&$0.524\pm0.743$&$1.88\pm0.86$\\
	$G_{SM}(T_{3L})$&$1.10\pm0.08$&$2.81\pm0.09$&$1.71\pm0.70$&$4.15\pm0.81$\\
	\hline
	\hline
	$M_{Z^{\prime}}=2.0$ TeV&$\Delta M_{t\bar t}<0.5$ TeV&$\Delta M_{t\bar t}<0.1$ TeV&$\Delta M_{t\bar t}<0.5$ TeV&$\Delta M_{t\bar t}<0.1$ TeV\\
	\hline
	$SM$&$0.010\pm0.167$&$0.011\pm0.216$&$0.02\pm1.86$&$0.03\pm2.84$\\
	$G_{LR}(LR)$&$0.745\pm0.153$&$2.40\pm0.16$&$1.13\pm1.67$&$3.70\pm1.82$\\
	$G_{LR}(R)$&$1.26\pm0.15$&$3.26\pm0.14$&$1.90\pm1.58$&$4.67\pm1.62$\\
	$G_{LR}(Y)$&$0.768\pm0.151$&$2.27\pm0.15$&$1.17\pm1.62$&$3.30\pm1.66$\\
	$G_{SM}(SM)$&$0.495\pm0.152$&$1.36\pm0.16$&$0.75\pm1.62$&$2.07\pm1.73$\\
	$G_{SM}(T_{3L})$&$1.54\pm0.14$&$3.12\pm0.15$&$2.33\pm1.51$&$4.54\pm1.67$\\
	\hline
	\end{tabular}
\caption{Summary of integrated $A_{RFB}$ values around the $Z^{\prime}$ peak for generalised models with $M_{Z^{\prime}}$=1.7 
and 2 TeV at the LHC at 14 and 8 TeV assuming 100 and 15 fb$^{-1}$ of integrated luminosity respectively.}\label{tab:LHC_GXX_ARFB}
\end{table}

\begin{landscape}
\begin{table}[h!]
	\centering
	\begin{tabular}{|c|c|c|c|c|c|c|c|c|c|c|c|c|}\hline
	$A_{L} \searrow A_{LL}$ & $SM$ & $E_{6}(\chi)$&$E_{6}(\eta)$&$E_{6}(\psi)$&$E_{6}(N)$&$E_{6}(S)$&$G_{LR}(B-L)$&$G_{LR}(LR)$&$G_{LR}(R)$&$G_{LR}(Y)$&$G_{SM}(SM)$&$G_{SM}(T_{3L})$\\ \hline
	$SM$         & -- & 62.6 & 5.1  & 10.6 & 38.7   & $>$300 & 77.5   & 3.8    & 2.7  & 3.9    & 4.0    & 3.0 \\
	$E_{6}(\chi)$&    & --   & 10.2 & 32.7 & $>$300 & 90.7   & $>$300 & 7.1    & 4.6  & 7.5    & 7.7    & 5.2 \\
	$E_{6}(\eta)$&    &      & --   & 48.4 & 13.1   & 5.6    & 9.5    & $>$300 & 45.3 & $>$300 & $>$300 & 70.7\\
	$E_{6}(\psi)$&    &      &      & --   & 51.7   & 12.3   & 28.2   & 24.2   & 11.1 & 26.8   & 28.2   & 13.7\\
	$E_{6}(N)$   &    &      &      &      & --     & 51.6   & $>$300 & 8.8    & 5.4  & 9.3    & 9.5    & 6.2\\
	$E_{6}(S)$   &    &      &      &      &        & --     & 117.8  & 4.2    & 3.0  & 4.3    & 4.4    & 3.3\\
	$G_{LR}(B-L)$ &    &      &      &      &        &        & --     & 6.6    & 4.3  & 6.9    & 7.1    & 4.9\\
	$G_{LR}(LR)$ & 0.8&      &      &      &        &        &        & --     & 95.9 & $>$300 & $>$300 & 200.0\\
	$G_{LR}(R)$  & 0.5&      &      &      &        &        &        & 9.3    & --   & 78.9   & 72.0   & $>$300 \\
	$G_{LR}(Y)$  & 0.9&      &      &      &        &        &        & 105.8  & 5.6  & --     & $>$300 & 151.7\\
	$G_{SM}(SM)$ & 1.6&      &      &      &        &        &        & 0.2    & 0.2  & 0.2    & --     & 133.9\\
     $G_{SM}(T_{3L})$& 0.5&      &      &      &        &        &        & 0.1    & 0.1  & 0.1    & 2.4    &  -- \\
	\hline
	\end{tabular}
\caption{Required integrated luminosity (in fb$^{-1}$) at the LHC at 14 TeV to ``distinguish'' $Z'$ models from the SM background and among themselves, for (upper triangle) $A_{LL}$ and (lower triangle) $A_L$, for $M_{Z^{\prime}}$=2.0 TeV. Models are disentangled if $s=3$. No value is given for $A_L$ for the $E_6$-type models because such asymmetry is too small to be observed.}\label{tab:Lumi_LHC14_ALL-AL}
\end{table}

\begin{table}[h!]
	\centering
	\begin{tabular}{|c|c|c|c|c|c|c|}\hline
	Spatial      &$SM$&$G_{LR}(LR)$ &$G_{LR}(R)$&$G_{LR}(Y)$&$G_{SM}(SM)$&$G_{SM}(T_{3L})$\\ \hline
	$SM$         & --         & 11.4(14.0)    & 5.6(6.6)      & 12.2(14.8)     & 35.7(42.6)    & 6.4(7.5)  \\
	$G_{LR}(LR)$ & 13.2(30.9) & --            & 55.0(59.1)    & $>$300($>$300) & 42.6(55.1)    & 83.5(94.0) \\
	$G_{LR}(R)$  & 7.1(14.3)  & 61.2(118.8)   & --            & 38.7(51.7)     & 11.3(13.4)    & $>$300($>$300) \\
	$G_{LR}(Y)$  & 13.7(33.0) & $>$300($>$300)& 55.3(98.4)    & --             & 52.3(63.0)    & 56.1(79.6) \\
	$G_{SM}(SM)$ & 40.0(97.6) & 40.9(122.5)   & 11.8(28.9)    & 44.4(144.5)    & --            & 14.0(16.6) \\
     $G_{SM}(T_{3L})$& 7.8(16.7)  & 94.0(204.9)   & $>$300($>$300)& 82.8(161.5)    & 14.0(37.5)    &  --    \\
	\hline
	\end{tabular}
\caption{Required integrated luminosity (in fb$^{-1}$) at the LHC at 14 TeV to ``distinguish'' $Z'$ models from the SM background and among themselves, for (upper triangle) $A_{RFB}(A_{OFB})$ and (lower triangle) $A_C(A_{F})$, for $M_{Z^{\prime}}$=2.0 TeV. Models are disentangled if $s=3$.}\label{tab:Lumi_LHC14_spatial}
\end{table}

\end{landscape}

\clearpage

\begin{landscape}
\begin{table}[h!]
	\centering
	\footnotesize
	\setlength{\tabcolsep}{4pt}
	\begin{tabular}{|c|c|c|c|c|c|c|c|c|c|c|c|c|}\hline
	$A_{LL}$ & $SM$ & $E_{6}(\chi)$&$E_{6}(\eta)$&$E_{6}(\psi)$&$E_{6}(N)$&$E_{6}(S)$&$G_{LR}(B-L)$&$G_{LR}(LR)$&$G_{LR}(R)$&$G_{LR}(Y)$&$G_{SM}(SM)$&$G_{SM}(T_{3L})$\\ \hline
   $SM$         & --       & 5.7(1.3) & 20.4(5.4) & 13.6(3.1) & 7.3(1.6) & 0.8(0.1) & 5.2(1.3)  & 23.6(8.8) & 30.7(12.9) & 24.9(9.1) & 23.3(10.3) & 29.6(18.0)\\
   $E_{6}(\chi)$& 3.8(0.9) & --       & 15.9(4.1) & 8.5(1.8)  & 1.6(0.3) & 4.9(1.1) & 0.6($\ll$ 1)& 19.4(7.5) & 27.0(11.6) & 20.6(7.8) & 19.1(9.1)  & 25.9(16.6)\\
   $E_{6}(\eta)$& 13.3(3.7)& 9.4(2.8) & --        & 7.4(2.3)  & 14.3(3.8)& 19.5(5.2)& 16.5(4.1) & 3.5(3.4)  & 9.7(7.5)   & 3.3(3.7)  & 3.2(4.9)   & 8.6(12.0)\\
   $E_{6}(\psi)$& 9.2(2.2) & 5.2(1.3) & 4.3(1.5)  & --        & 6.8(1.6) & 12.7(3.0)& 9.1(1.8)  & 11.0(5.7) & 17.8(9.8)  & 11.4(5.9) & 10.6(7.2)  & 16.6(14.5)\\
   $E_{6}(N)$   & 4.8(1.0) & 1.0(0.2) & 8.3(2.7)  & 4.2(1.2)  & --       & 6.4(1.4) & 2.2(0.3)  & 17.8(7.2) & 25.2(11.3) & 18.8(7.5) & 17.4(8.8)  & 24.1(16.2)\\
   $E_{6}(S)$   & 0.6(0.1) & 3.1(0.8) & 12.6(3.6) & 8.5(2.1)  & 4.2(0.9) & --       & 4.3(1.1)  & 22.8(8.6) & 29.8(12.7) & 23.9(8.9) & 22.5(10.2) & 28.7(17.8)\\
   $G_{LR}(B-L)$ & 3.4(0.9) &0.4($\ll$ 1)& 9.8(2.8)  & 5.7(1.3)  & 1.4(0.2) & 2.8(0.8) & --        & 20.0(7.5) & 27.7(11.6) & 21.3(7.8) & 19.7(9.1)  & 26.5(16.6)\\
   $G_{LR}(LR)$ & 15.4(6.2)&11.3(5.3) & 1.6(2.3)  & 6.1(3.9)  & 10.1(5.1) &14.7(6.1) & 11.7(5.3) & --        & 5.9(4.1)   & 0.5(0.3)  & 0.4(1.6)   & 4.7(8.3)\\
   $G_{LR}(R)$  & 18.2(9.0)& 14.0(8.1)& 4.5(5.1)  & 9.0(6.6)  & 12.9(7.9)& 17.4(8.9)& 14.4(8.1) & 3.1(2.9)  & --         & 7.1(3.8)  & 6.3(2.5)   & 1.3(3.7) \\
   $G_{LR}(Y)$  & 15.1(6.5)& 11.0(5.6)& 1.3(2.6)  & 5.8(4.2)  & 9.9(5.4) & 14.4(6.4)& 11.4(5.6) & 0.3(0.3)  & 3.4(2.7)   & --        & 0.1(1.3)   & 5.8(8.0)\\
   $G_{SM}(SM)$ & 15.0(7.2)& 10.8(6.3)& 1.1(3.3)  & 5.6(4.9)  & 9.7(6.1) & 14.3(7.1)& 11.3(6.3) & 0.5(1.1)  & 3.5(1.9)   & 0.2(0.8)  & --         & 5.1(6.6)\\
$G_{SM}(T_{3L})$&17.3(11.0)&13.2(10.1)& 3.6(7.1)  & 8.1(8.7)  &12.0(10.0)&16.6(11.0)& 13.6(10.1)& 2.1(5.1)  & 0.9(2.2)   & 2.4(4.9)  & 2.6(4.1)   &  -- \\
	\hline
	\end{tabular}
\caption{Significance for $A_{LL}$ as in Tables~\ref{tab:LHC_E_6_ALL}--\ref{tab:LHC_GXX_ALL}, for the LHC at 14 TeV only. Upper triangle for $M_{Z'}=1.7$ TeV and lower triangle for $M_{Z^{\prime}}$=2.0 TeV. Figures refer to $\Delta M_{t\bar t}<100$(500) GeV.}\label{tab:signif_LHC14_ALL}
\end{table}

\begin{table}[h!]
	\centering
	\begin{tabular}{|c|c|c|c|c|c|c|}\hline
	$A_L$        &$SM$&$G_{LR}(LR)$ &$G_{LR}(R)$&$G_{LR}(Y)$&$G_{SM}(SM)$&$G_{SM}(T_{3L})$\\ \hline
	$SM$         & --         & 50.0(22.8) & 63.9(37.3) & 45.9(22.0) & 35.0(19.2)  & 61.8(47.4)\\
	$G_{LR}(LR)$ & 34.1(12.2) & --         & 15.3(9.3)  & 4.5(0.6)   & 93.5(29.9)  & 123.0(49.5) \\
	$G_{LR}(R)$  & 42.8(18.5) & 9.8(7.2)   & --         & 19.8(9.9)  & 108.8(39.9) & 123.0(60.3) \\
	$G_{LR}(Y)$  & 31.5(11.8) & 2.9(0.4)   & 12.7(7.6)  & --         & 89.0(29.3)  & 118.5(48.9) \\
	$G_{SM}(SM)$ & 23.6(10.1) & 65.3(25.3) & 75.1(32.4) & 62.4(24.8) & --          & 29.5(18.9) \\
     $G_{SM}(T_{3L})$& 40.8(22.8) & 84.8(39.7) & 84.8(46.9) & 81.8(39.3) & 19.4(14.4)  &  --    \\
	\hline
	\end{tabular}
\caption{Significance for $A_{L}$ as in Table~\ref{tab:LHC_GXX_AL}, for the LHC at 14 TeV only. Upper triangle for $M_{Z'}=1.7$ TeV and lower triangle for $M_{Z^{\prime}}$=2.0 TeV. Figures refer to $\Delta M_{t\bar t}<100$(500) GeV.}\label{tab:signif_LHC14_AL}
\end{table}
\end{landscape}

\begin{table}[h!]
	\centering
	\begin{tabular}{|c|c|c|c|c|c|c|}\hline
	$A_{RFB}$      &$SM$&$G_{LR}(LR)$ &$G_{LR}(R)$&$G_{LR}(Y)$&$G_{SM}(SM)$&$G_{SM}(T_{3L})$\\ \hline
	$SM$         & --       & 13.6(4.0) & 19.0(7.2) & 13.1(4.2) & 7.8(2.7) & 18.5(9.1)  \\
	$G_{LR}(LR)$ & 8.9(3.2) & --        & 6.4(3.2)  & 0.6(0.2)  & 6.9(1.4) & 5.8(5.2) \\
	$G_{LR}(R)$  & 12.6(5.6)& 4.0(2.4)  & --        & 7.1(3.0)  & 13.4(4.6)& 0.6(2.0) \\
	$G_{LR}(Y)$  & 8.6(3.4) & 0.6(0.1)  & 4.8(2.3)  & --        & 6.3(1.6) & 6.4(5.0) \\
	$G_{SM}(SM)$ & 5.0(2.1) & 4.6(1.2)  & 8.9(3.6)  & 4.1(1.3)  & --       & 12.7(6.6) \\
     $G_{SM}(T_{3L})$& 11.8(7.0)& 3.3(3.8)  & 0.7(1.4)  & 4.0(3.7)  & 8.0(5.1) &  --    \\
	\hline
	\end{tabular}
\caption{Significance for $A_{RFB}$ as in Table~\ref{tab:LHC_GXX_ARFB}, for the LHC at 14 TeV only. Upper triangle for $M_{Z'}=1.7$ TeV and lower triangle for $M_{Z^{\prime}}$=2.0 TeV. Figures refer to $\Delta M_{t\bar t}<100$(500) GeV.}\label{tab:signif_LHC14_ARFB}
\end{table}

\section*{Acknowledgements}
The work of KM and SM is partially supported through the NExT Institute. LB is supported by the 
Deutsche Forschungsgemeinschaft through the Research Training Group grant
GRK\,1102 \textit{Physics of Hadron Accelerators}. We would like to thank E. Alvarez for pointing out the discussion on systematic uncertainties.

\newpage

\end{document}